\documentclass[smallcondensed,numbers]{svjour3}

\usepackage{graphicx}
\usepackage{amsmath,amssymb} % define this before the line numbering.
\usepackage{color}

\usepackage{times}
\usepackage{epsfig}
\usepackage{subfigure}
\usepackage{array}
\usepackage{multirow}
\usepackage{natbib}
\usepackage{dsfont}
\usepackage[export]{adjustbox}
\usepackage{url}
\usepackage{setspace}

\usepackage[font=doublespacing]{caption}

\usepackage{pdflscape}

\newcolumntype{V}{>{$\vcenter\bgroup\hbox\bgroup}c<{\egroup\egroup$}}

\graphicspath{{../../../2014/2014_ASONAM/imgs/}{../imgs-new/}}

\begin{document}

\title{Using Twitter to Learn about the Autism Community}

\titlerunning{Using Twitter to Learn about the Autism Community}

\author{Adham Beykikhoshk, Ognjen~Arandjelovi\'c, Dinh~Phung, Svetha~Venkatesh, and~Terry Caelli}

\institute{
Centre for Pattern Recognition and Data Analytics (PRaDA)\\
School of Information Technology\\
Deaking University\\
Waurn Ponds
Geelong 3216\\
VIC\\
Australia}

\maketitle

\doublespacing

\begin{abstract}
Considering the raising socio-economic burden of autism spectrum disorder (ASD), timely and evidence-driven public policy decision making and communication of the latest guidelines pertaining to the treatment and management of the disorder is crucial. Yet evidence suggests that policy makers and medical practitioners do not always have a good understanding of the practices and relevant beliefs of ASD-afflicted individuals' carers who often follow questionable recommendations and adopt advice poorly supported by scientific data. The key goal of the present work is to explore the idea that Twitter, as a highly popular platform for information exchange, could be used as a data-mining source to learn about the population affected by ASD -- their behaviour, concerns, needs etc. To this end, using a large data set of over 11 million harvested tweets as the basis for our investigation, we describe a series of experiments which examine a range of linguistic and semantic aspects of messages posted by individuals interested in ASD. Our findings, the first of their nature in the published scientific literature, strongly motivate additional research on this topic and present a methodological basis for further work.\\~\\
%The autism spectrum disorder (ASD) is increasingly being recognized as a major public health issue which affects approximately 0.5-0.6\% of the population. Promoting the general awareness of the disorder, increasing engagement with the affected individuals and their carers, and understanding the success of penetration of the current clinical recommendations in the target communities, is crucial in driving research as well as policy. The aim of the present work is to investigate if Twitter, as a highly popular platform for information exchange, can be used as a data-mining source which could aid in the aforementioned challenges. Specifically, using a large data set of harvested tweets, we present a series of experiments which examine a range of linguistic and semantic aspects of messages posted by individuals interested in ASD. Our findings, the first of their nature in the published scientific literature, strongly motivate additional research on this topic and present a methodological basis for further work.\\~\\
\end{abstract}

\section{Introduction}
In this paper we are interested in leveraging the remarkable increase in the use of social media, and Twitter in particular, to  obtain information about specific demographic communities which are difficult to reach by conventional means. To illustrate our ideas and demonstrate their effectiveness, herein we focus on the population affected by the autism spectrum disorder (ASD) -- a neurodevelopmental disorder which has been attracting increasing attention as much for its complex and varied aetiology, as for the associated and rapidly growing socioeconomic burden. The population of interest in this work includes both individuals who have been diagnosed with ASD themselves, as well as those who are indirectly but significantly impacted such as the family members and carers of ASD sufferers. To the best of our knowledge ours is the first work of this nature. The present paper builds on the preliminary results reported in~\cite{BeykAranPhunVenk+2014} and describes a number of novel findings and experiments whose results further support the key underlying idea and motivate additional work in this direction. Most notable novelties include:
\begin{itemize}
  \item additional statistics of the collected tweet corpus,\\[-8pt]
  \item quantitative analysis of Zipf's law for tweets and ASD-related tweets in particular,\\[-8pt]
  \item examination of statistical significance of differential part-of-speech characteristics,\\[-8pt]
  \item experiments using least absolute shrinkage and selection operator (LASSO),\\[-8pt]
  \item additional evidence of content saliency using an Alzheimer's disease tweet corpus, and \\[-8pt]
  \item new analysis of the procedure for the selection of the bootstrap keyword set.\\
\end{itemize}
It should also be noted that the results reported in this paper include a significantly expanded corpus of tweets (nearly an order of magnitude greater) as we continued collecting data following the publication of our original work.

The remainder of this paper is structured as follows. In the next section we describe some of the key aspects of the autism spectrum disorder which motivated our focus on this particular condition. In Section~\ref{s:pref} we review relevant previous work on the use of Twitter for data mining and highlight the key techniques and methods in the field. Section~\ref{s:methodsRes} describes the main contribution of our work -- we start by describing the manner in which the data used for the present study was collected and follow up with a description, results, and  discussion of a sequence of increasingly complex experiments. A summary of the work and its findings is made in Section~\ref{s:conc}.

\section{Motivation and relevant background}
Autism spectrum disorder is a life-long neurodevelopmental disorder with poorly understood causes on the one hand, and a wide range of potential treatments supported by little evidence on the other. The disorder is characterized by severe impairments in social interaction, communication, and in some cases cognitive abilities, and typically begins in infancy or at the very latest by the age of three. ASD is recognized as comprising an aetiologically and clinically heterogeneous group of conditions whose diagnosis remains to be based solely on the complex behavioural phenotype~\cite{Mile2011}. According to the definition in the latest version (5th edition) of the Diagnostic and Statistical Manual of Mental Disorders, the autism spectrum disorder includes disorders which were previously diagnosed with more specificity as autism, Asperger syndrome, Rett syndrome, childhood disintegrative disorder, and `pervasive developmental disorder not otherwise specified'~\cite{DSM5}. Current evidence suggests that approximately 0.5-0.6\% of the population is afflicted by ASD though the actual diagnosis rate is on the increase due to the broadening diagnostic criteria~\cite{BaxtBrugErskSche+2015}. The condition is usually detected in early childhood when an abnormal lack of social reciprocity is observed.

\begin{figure}[htbp]
  \centering
  \includegraphics[width=0.99\linewidth]{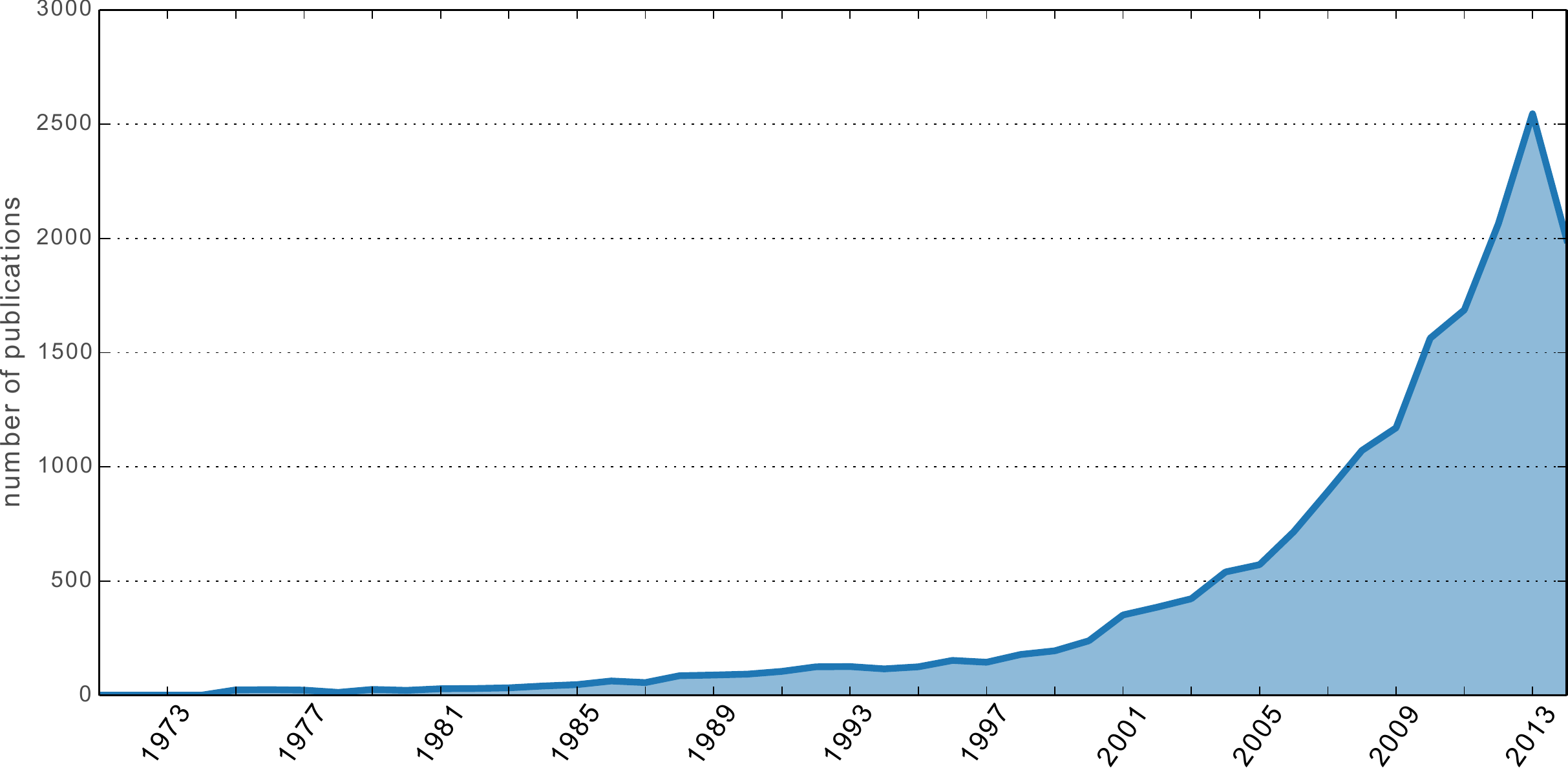}
  \caption{The number of academic papers in English with the word ``autism'' in the title or abstract as indexed by the PubMed portal which interfaces the US National Library of Medicine life and biomedical sciences database. The earliest work is that by Kanner in 1946~\cite{Kann1946}; thereafter an exponential increase is readily observed. Note that the count for 2014 is incomplete as it includes papers indexed by PubMed only up to and including May 2014. }
  \label{f:nPapers}
\end{figure}

Considering the social and economic burden of ASD it is unsurprising that it has been attracting an increasing amount of research attention; numerous longitudinal, epidemiological, and family studies have been conducted~\cite{BeykPhunAranVenk2015}. This is illustrated by the plot in Fig.~\ref{f:nPapers} which shows the number of academic papers in English with the word ``autism'' in the title or abstract (as indexed by the PubMed portal which interfaces the US National Library of Medicine life and biomedical sciences database). The earliest work is that by Kanner in 1946~\cite{Kann1946} with a readily observed exponential increase thenceforth. Note that the count for the year 2014, which seemingly bucks the trend, is in fact perfectly in agreement with the overall increase as only papers indexed by PubMed up to and including May 2014 are included.

Although the last few decades have seen significant progress in the study of ASD, the still relatively poorly understood aetiology of the condition, its phenotypical heterogeneity~\cite{LevyMandSchu2009}, and stigma associated with mental conditions~\cite{Gray1993}, have all contributed to the penetration of beliefs, and behavioural and educational interventions which are often questionable~\cite{WarrMcPhSathFoss2011} and poorly supported by evidence (e.g.\ gluten-free and casein-free diets, and cognitive behavioural therapy~\cite{DaniWood2013}), and sometimes outright in conflict with science~\cite{TremBalaRoss2005}. For example a recent review of early intensive behavioural and developmental interventions for young children with ASD found 1 existing study as being of good quality, 10 as fair quality, and 23 as poor quality~\cite{WarrMcPhSathFoss2011}. From the public policy point of view, understanding the practices and beliefs of parents and carers of ASD-affected individuals is crucial, yet often lacking~\cite{HarrRoseGarnPatr2006}.

Thus the key idea motivating the present work is that the rapid rise in the adoption of social media as a platform for the expression and exchange of ideas, which facilitates the emergence of special interest communities, can be used to study and monitor the beliefs and practices of the population affected by ASD. Considering the challenge of reaching and engaging with this specific target population, our findings pave the way for further work of potential significant benefit to public health. These benefits include the enrichment of the corpus of knowledge of the condition itself by the medical community, and the increased understanding of the practices and concerns of those affected by ASD.

\section{Previous work}\label{s:pref}
While this is the first work exploring the possibility of using Twitter data for the extraction of ASD-related information, the broad idea of data-mining Twitter is not new and has been employed successfully in a variety of applications. At the same time it is important to stress at the very outset that most of the previous work on this topic has not been automatic, that is, analysis was performed `manually' by humans. This is a laborious process which severely limits how much data can be processed. Additionally, the use of human intelligence rather than computer-based methods means that in the case of many reported results, it is not clear that the same results could be obtained automatically due to a possible semantic gap.

A popular research direction focuses on various forms of prediction based on tweet sentiments~\cite{JianYuZhouLiu+2011} inferred from the tweet's emoticons \cite{BifeFran2010} or using linguistics-based classifiers~\cite{AgarXieVovsPass2011}. For example, Asur and Huberman~\cite{AsurHube2010} showed that tweet posting rate can be used to forecast film box office revenues, while Baucom \textit{et al.}~\cite{BaucSanjLiuChen2013} used sentiment to analyse the relationship between the content of tweets and the outcomes of NBA Playoff games. Bollen \textit{et al.} \cite{BollMaoPepe2011} showed that tweet sentiments, aggregately seen as a reflection of public mood levels, can be used to predict the incidence of socio-political, cultural, and economic events. Mitchell \textit{et al.} \cite{MitcFranHarrDodd+2013} used geo-tagged tweets across the United States to estimate localized changes in a variety of sociometric indices such as happiness, the level of education, and obesity rates.

The quasi-realtime nature of Twitter also makes it a potentially valuable resource for the detection and management of emergency situations \cite{VermViewCorvPale+2011}. For example, Robinson \textit{et al.} \cite{RobiPoweCame2013} described an earthquake detection system, while Sakaki \textit{et al.} \cite{SakaOkazMats2010} used a spatio-temporal model of tweet frequencies to infer the location of the epicentre of an earthquake.

Closer in spirit to the nature of work in the present paper is the corpus of work on the use of Twitter in the domain of health care. For example, Paul and Dredze \cite{PaulDred2012} used tweets to extract words related to symptoms and treatments, and a topic model to associate them with the corresponding ailments. In their subsequent work \cite{PaulDred2011}, the model was extended to track the spread of illnesses over time, measure behavioural risk factors, and analyse symptoms and medication usage. The use of LDA-based health topic modelling was explored by Prier \textit{et al.} \cite{PrieSmitGiraHans2011}. Dementia-related tweets were the focus of work by Robillard \textit{et al.} \cite{RobiJohnHennBeat+2013} who collected relevant tweets over a 24h period using keyword filtering and used them to discover the dominant chatter themes. Similarly, Scanfeld \textit{et al.} \cite{ScanScanLars2010} used Twitter to analyse the patterns of antibiotic use. Jashnisky \textit{et al.} \cite{JashBurtHansWest+2014} studied the relationship between Twitter conversations deemed to reflect high suicide risk and actual suicide rates in the United States. They demonstrated that high risk individuals may be recognized from their social media status. Lastly, Himelboim and Han \cite{HimeHan2014} examined the connectivity patterns of Twitter users interacting within a specific online community of cancer-affected individuals.

In contrast to the sporadic studies on different diseases described above, the use of Twitter data in the management of highly contagious diseases like influenza has attracted a more concentrated research effort. For example, Culotta \cite{Culo2010} investigated whether the frequency of influenza epidemic-related tweets can be related to `ground truth' data from centres for disease control and prediction. Achrekar \textit{et al.} \cite{HarsGandLazaYu+2011} showed that the emergence and the spread of epidemic influenza can be predicted and tracked from the location and demographic information of users of relevant tweets. A similar approach was also described by Li and Cardie \cite{LiCard2013}. Yet further evidence of the power of Twitter data was presented by Chew and Eysenbach~\cite{ChewEyse2010} who demonstrated that the spatio-temporal distribution of relevant tweet frequencies during the 2009 H1N1 outbreak closely matches the disease spreading pattern.

Although the use of Twitter for data-mining information related to ASD has not been explored yet, there has been some preliminary work on the use of other social media and ASD. For example, Newton \textit{et al.} \cite{NewtKramMcIn2009} used Linguistic Inquiry and Word Count (LIWC) dictionaries to compare writing patterns of individuals with ASD and those of neuro-typical bloggers.

\section{Methods, results, and discussion}\label{s:methodsRes}
Having outlined the motivation and our ultimate vision for this work, and placed it in context of previous research on data-mining social media, we now turn our attention to the main contribution of the present paper. We start by describing the data set we used for our analysis.

\subsection{Data acquisition}\label{ss:data}
Twitter's Terms of Service explicitly prohibit the sharing or redistribution of tweets, including for research purposes. Consequently, there was no public data set that we could use as a standard benchmark in this study. Instead, we collected a large data set ourselves.

Twitter API offers different means of retrieving tweets. In particular, we used its `search' and 'streaming' functions. The former allows the retrieval of historical tweets based on the presence of specific keywords and meta-data constraints (e.g.\ on the language or user location). After being posted, a tweet can be obtained in this manner for up to a week. The streaming API allows a quasi-realtime retrieval of tweets as they are posted, retrieving a sample of approximately 1\% of all tweets.  The search API was most valuable for us for collecting ASD-related content, as we will describe in detail shortly. Conversely, the streaming API allowed us to obtain `control' data, unrelated to ASD, since this set could not be well characterized \textit{a priori} using a compact set of keywords.

We collected only tweets posted in English. To facilitate a comparative analysis, we collected two non-overlapping data sets. The first of these, which we will refer to as the ASD subset, comprises tweets which concern ASD. Specifically, we defined the ASD subset as comprising tweets which contain any of the four keywords ``autism'', ``adhd'', ``asperger'', and ``aspie'' (or any of their derivatives obtained by suffixation), and the control subset as comprising all other tweets. In total this resulted in a corpus of 5,650,989 ASD-related tweets collected in the period starting on 26 August 2013 and ending on 1 Oct 2014 (i.e.\ more than 13 consecutive months). Of these, 3,493,742 were original tweets which were produced by 1,771,274 unique users, with the remaining 28\% of the messages being re-tweets. Approximately 25\% of the collected tweets (2,260,284) contain so-called hashtags (see Section~\ref{sss:Hash} for discussion and analysis). Although this information was not used in the present work we also report that in our corpus 70,925 tweets are geo-tagged (it should be noted that geo-tags, being controlled entirely by the user, are not necessarily correct location identifiers), 2,599,395 contain URLs, 464,190 were sent in reply, and 3,330,096 mention another user.

It is important to observe that it is \emph{not} our claim that tweets in the ASD subset were necessarily posted by individuals suffering from ASD. While some of the messages in this subset do fall into this category, the subset will also include posts by individuals affected by ASD in a looser sense, e.g.\ parents or carers of those who suffer from ASD, or indeed medical professionals interested in the condition.

\begin{figure}[htbp]
  \centering
  \includegraphics[width=0.9\linewidth]{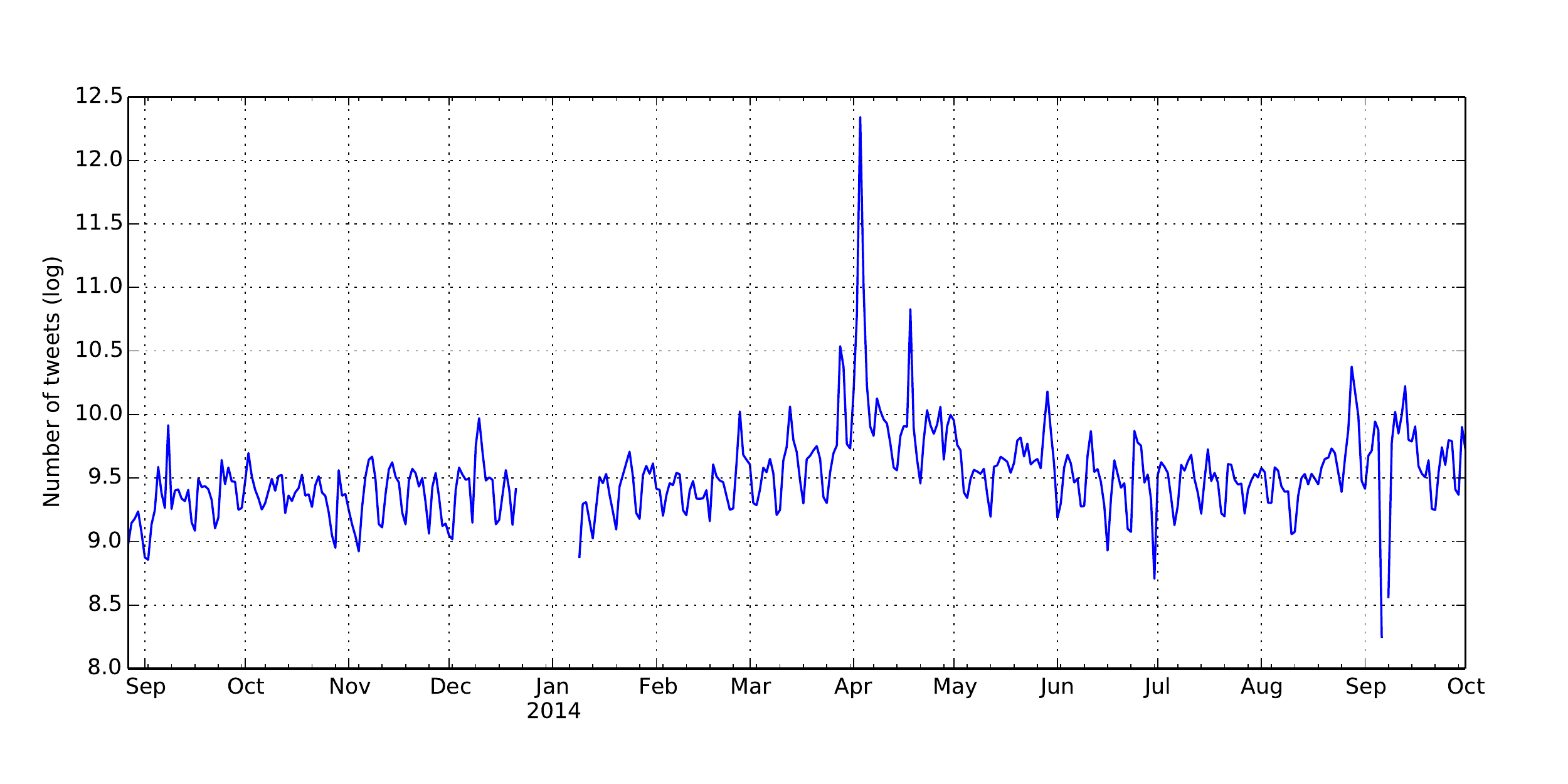}
  \caption{Number of ASD-related tweets we collected each day over the course of our data acquisition period; notice weekly periodicity. To make the usual day to day variability visible in the presence of the sharp peak on 2 April (World Autism Awareness Day) logarithmic scaling was used for ordinate values. The approximately three week gap in data collection during the 2014-2015 holiday period was caused by the failure of our computer system. }
  \label{f:nTweets}
\end{figure}

\subsubsection{Pre-processing}\label{sss:Preprocess}
Much of the work in the present paper concerns the analysis of topics discussed by means of Twitter. In this context it is beneficial to have different inflections of the same word normalized and represented by a single term. In linguistics this process is referred to as lemmatization and we apply it automatically using the freely available Natural Language Toolkit (NLTK)~\cite{Perk2010}. In addition, we remove the so-called `stop words' which do not carry much meaning themselves (e.g.\ articles and connectives), as well as all punctuation marks and emoticons. To illustrate the effects of our pre-processing we present a few examples.\\~\\
\textbf{Original tweet:}~
\begin{quote}\it
  Looks like we will have more \#autism research happening for children in \#EarlyIntervention next year! :-) \#VisualSupports \#MobileTech
\end{quote}
\textbf{Pre-processed result:}~
\begin{quote}\it
  look like autism research happen child earlyintervention visualsupports mobiletech
\end{quote}
\textbf{Original tweet:}~
\begin{quote}\it
  Authors who see autism as ``tremendously burdening'' elicit dire views of autism from parents http://j.mp/1FyqXwF  ``Ethical approval: none''
\end{quote}
\textbf{Pre-processed result:}~
\begin{quote}\it
  author see autism tremend burden elicit dire view autism parent url ethic approv none
\end{quote}
\textbf{Original tweet:}~
\begin{quote}\it
  101 autism: Genetic analysis of individuals with autism finds gene deletions - Using powerful genetic sequencing. http://is.gd/UhprQK
\end{quote}
\textbf{Pre-processed result:}~
\begin{quote}\it
  101 autism genet analysi individu autism find gene delet use power genet sequenc  url
\end{quote}
\textbf{Original tweet:}~
\begin{quote}\it
  \#Apple \#Censorship \& Dr. Brian Hooker Interview exposing CDC Cover-up of the Vaccine \& Autism Link on .@rediceradio  http://youtu.be/19uvPtg6SPI
\end{quote}
\textbf{Pre-processed result:}~
\begin{quote}\it
  appl censorship dr brian hooker interview expos cdc cover vaccin autism link atus url
\end{quote}

\subsection{Methods and results}\label{ss:MethodsRes}
In this section we describe a series of experiments aimed at discovering the properties of the collected data. We start with a generic quantitative linguistic analysis and proceed with increasingly domain-specific considerations which examine tweet content and the potential differentiation between the ASD-related and the control corpora.

\subsubsection{Power law: Zipf's distribution}\label{sss:Zipf}
The first experiment we conducted was set to find out if tweets, both in the ASD corpus as well as the control group, obey the so-called Zipf's law. In its general form, this empirical law posits that the frequency $P_i$ of an `event' is approximately a power function of its frequency rank $r_i$, i.e.\ $P_i \propto r_i^a$ where $r_i$ is the rank ($r_i=1,\ldots$) of the event when events are ordered by their frequency of occurrence from high to low and $a<-1$. Zipf's distribution can be seen to be a type of a power law probability distribution. This statistical regularity is observed in a variety of domains ranging from deadly quarrel~\cite{Rich1948} and wealth distribution analysis~\cite{BoucMeza2000}, to information retrieval~\cite{ClauShalNewm2009} and quantitative linguistics~\cite{ClauShalNewm2009}. In quantitative linguistics, an event corresponds to the usage of a particular word or, more precisely in the context of the present paper, a particular term resulting from the pre-processing of a word as described in Section~\ref{sss:Preprocess}.

Evidence from previous work suggests that conventional texts, such as newspapers or books, do result in Zipfian distributions of word frequencies~\cite{ClauShalNewm2009}. However, it is not clear from this that the same applies to tweets. Firstly, tweets are restricted in length to 140 characters which by itself may alter linguistic characteristics of posted messages. In addition, the nature of Twitter as a communication medium invariably introduces a self-selecting aspect: neither can the corpus of Twitter users be considered to be a random sample from the population, nor can the memetic content of tweets be expected to match that of texts examined by previous work, such as newspapers and books.

\begin{figure}[htbp]
  \centering
  \subfigure[]{\includegraphics[width=0.9\linewidth]{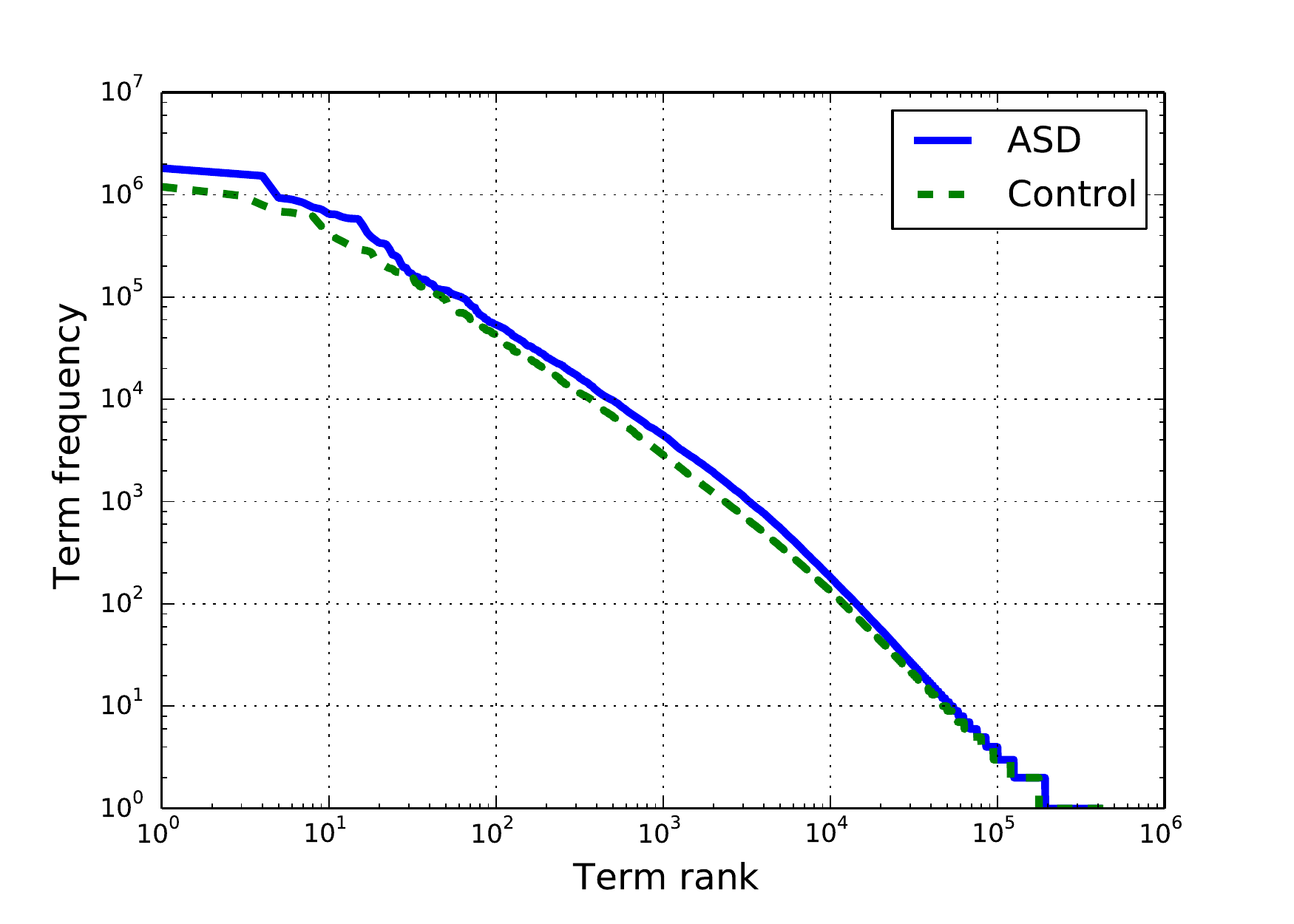}\label{f:ZipfDist}}
  \subfigure[]{\includegraphics[width=0.9\linewidth]{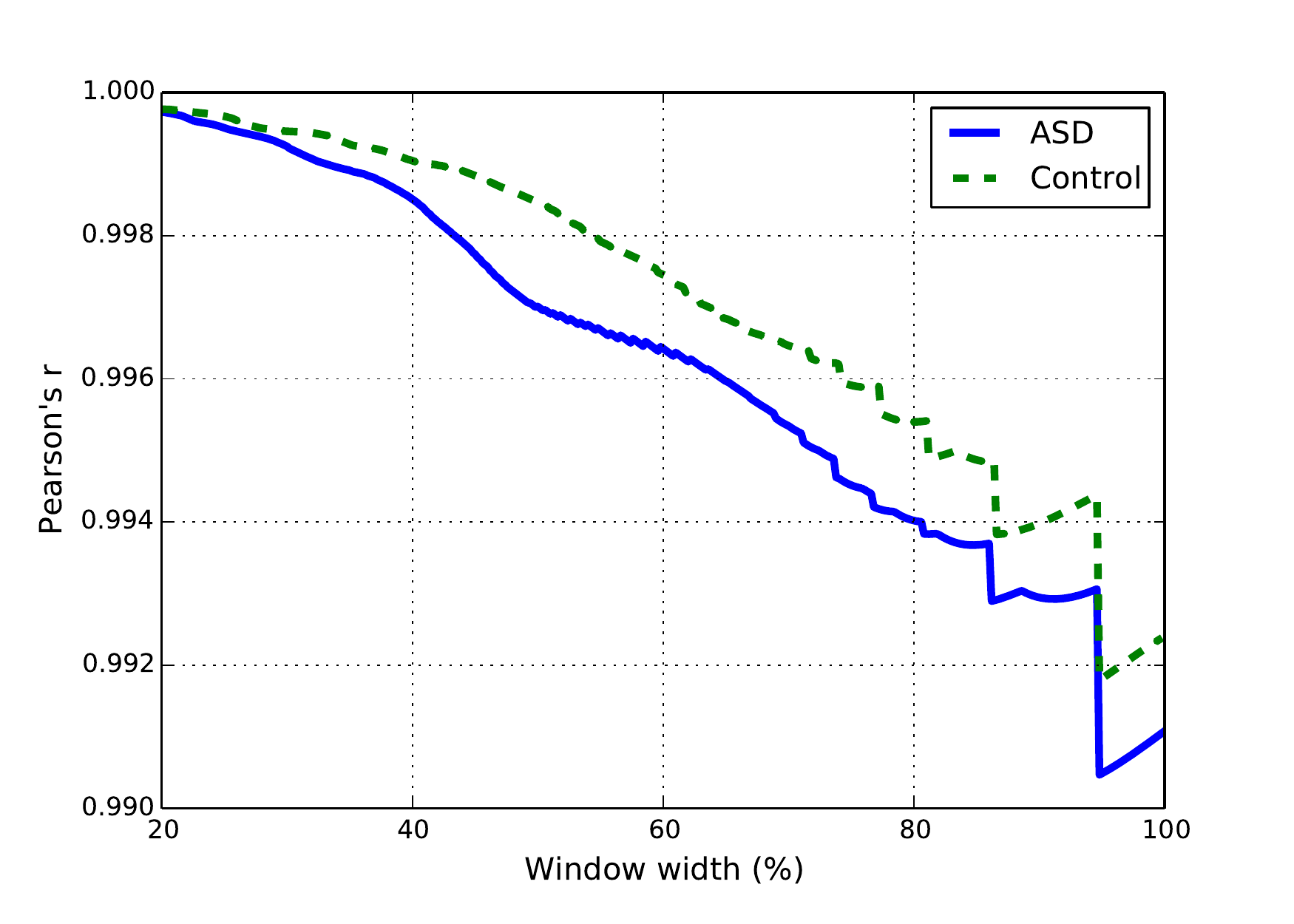}\label{f:rDist}}
  \caption{ (a) A log-log plot showing the dependence of tweet term (word) frequency on its frequency rank within the corpus of all terms in our data set. Approximately linear dependency is exhibited by data extracted from both the ASD and the control data sets, in agreement with the Zipfian power law distribution. (b) Linear dependency is quantified using the Pearson's $r$ statistic. To account for the expected artefacts at the ends of the distribution, the statistic is plotted as a function of the width of the window centred at the logarithmically median rank and spanning 20--100\% of the logarithmic range of possible ranks. Pearson's $r$ attains values of over 0.995 for the central portion of the distribution, and drops slightly (to approximately 0.992) for the entire range when the end artefacts are included. }
\end{figure}

In our data set, the ASD corpus of tweets contains 47,048,097 terms (pre-processed words) of which 402,946 are unique; the control subgroup used 32,846,321 terms of which 528,755 are unique. The key findings are summarized in Fig.~\ref{f:ZipfDist} which shows the plot of tweet term frequency as a function of its overall term frequency rank. It is readily apparent that the characteristics of both the ASD and the control group are nearly identical. What is more, both can be seen to exhibit approximately linear behaviour on this plot; a mild deviation from this behaviour can be observed for the most frequent terms. There are two main reasons for this. Firstly, it is generally the case that the simple functional form of Zipf's law fails to be observed for the top-ranking events. Secondly, our use of the logarithmic scale for the abscissa means that the left-hand tail of the graph is sparsely populated by data points which increases the corresponding error margins in this plot.

To quantify the linearity of dependence plotted in Fig.~\ref{f:ZipfDist} we used Pearson's $r$ statistic. Specifically, we computed the value of the statistic first for the window which spans 20\% of the ranks on the logarithmic scale and is centred at the logarithmic median of the ranks (i.e.\ the centre of the plot in Fig.~\ref{f:ZipfDist}), and then proceeded to increase the width recomputing the value for the encompassed range until all ranks were included. The variation of Pearson's $r$ as a function of the window width is plotted in Fig.~\ref{f:rDist} confirming our previous observations. The linearity of the central portion of the distribution is nearly perfect as witnessed by the $r$ value of 0.999. As more data is added a slight non-linearity is observed, the statistic dropping down to approximately 0.992 for the entire range.

In addition to their novelty, these findings are interesting in terms of directing our research towards our ultimate goal of automatically retrieving and analysing the content of ASD-related tweets. In particular, the discovery that tweets, including those in the sub-group of our primary interest, obey Zipf's law suggests that it is sensible to adopt and explore the use of a broad range of well-known and well-understood text representations and methods of analysis. Indeed we do this next.

\subsection{Message length}\label{ss:Length}
\begin{figure}[htb]
  \centering
  \includegraphics[width=0.9\linewidth]{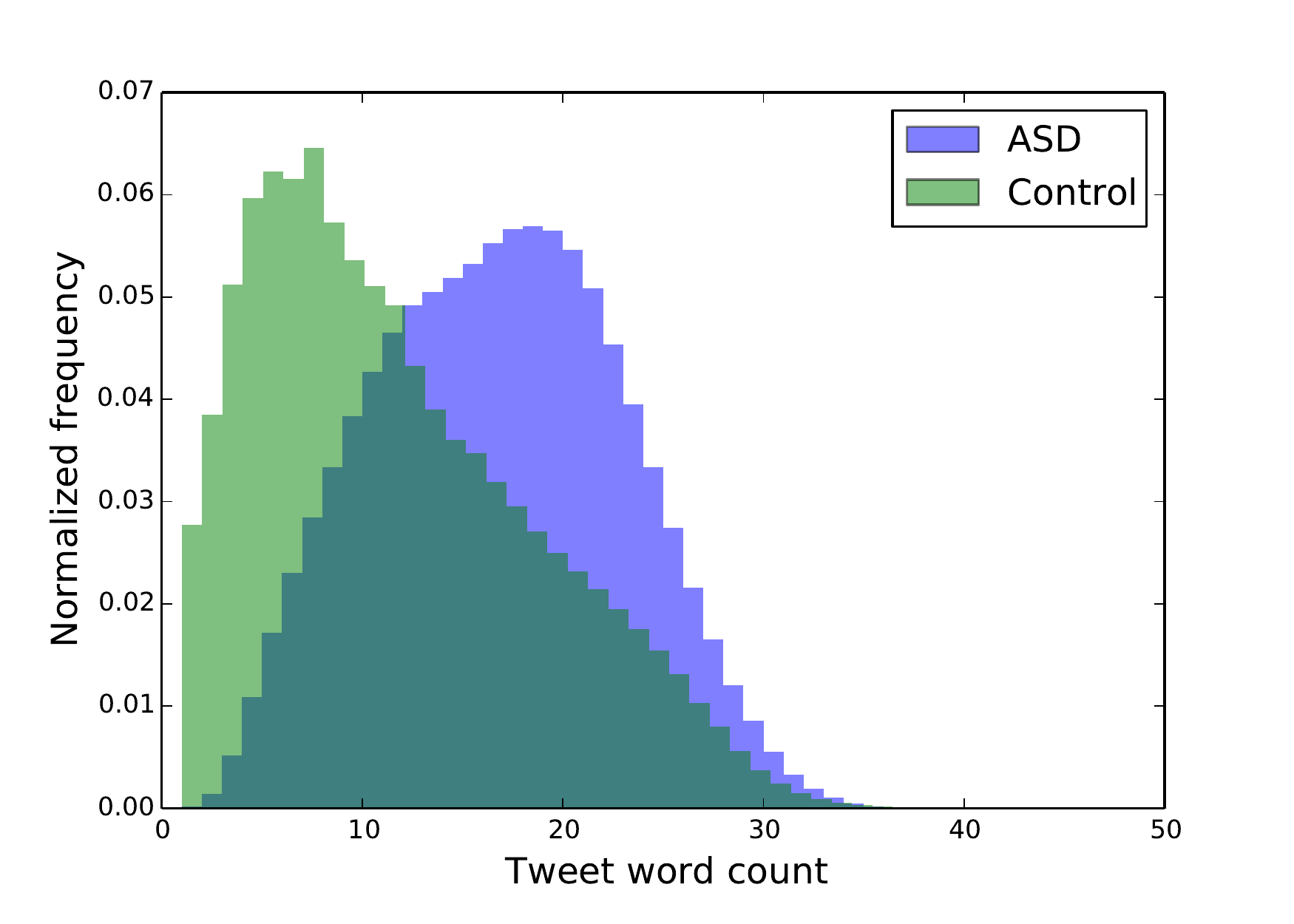}
  \caption{Normalized histograms of tweet word counts for the two subsets. As expected, log-normal distributions are observed but, importantly in the context of the present work, with different parameters (most obviously the means, $p<0.01$).}
  \label{f:TweetLen}
\end{figure}

Having established that Twitter messages conform to some of the same general linguistic rules as conventional texts do, our next goal was to explore any differential characteristics exhibited by our ASD and control groups. Recall from Section~\ref{ss:data} that our data set is balanced in the number of tweets, that is, the number of tweets in the ASD corpus is the same as in the control corpus. Yet, as pointed out in the previous section, the term counts in the two data sets are different, respectively 47,048,097 and 32,846,321 -- a significant difference of approximately 43\%. It is a direct implication of this observation that the average term count per tweet is greater in the ASD group, which is the first indication of there being a difference between tweets in this group and the remainder of our data corpus. The tendency of this group to post longer messages is further corroborated by comparing the corresponding histograms of tweet word counts, which are shown in Fig.~\ref{f:TweetLen}. While the tweet word count in both sub-groups can be seen to be log-normally distributed, the two distributions have substantially different means ($p<0.01$).

\subsection{Content analysis}\label{ss:Content}
Our next aim was to explore if the collected data offers evidence that tweets in the ASD and control data sets differ significantly by their content (i.e.\ topic of conversation) and if so, if automatic methods could be employed in the analysis of this content.

\subsubsection{Word frequency}\label{sss:WordFreq}
\begin{figure*}[htb]
  \centering
  \subfigure[ASD `word-cloud']{\includegraphics[bb=700bp 205bp 1900bp 2900bp,clip,angle=-90,width=0.9\textwidth]{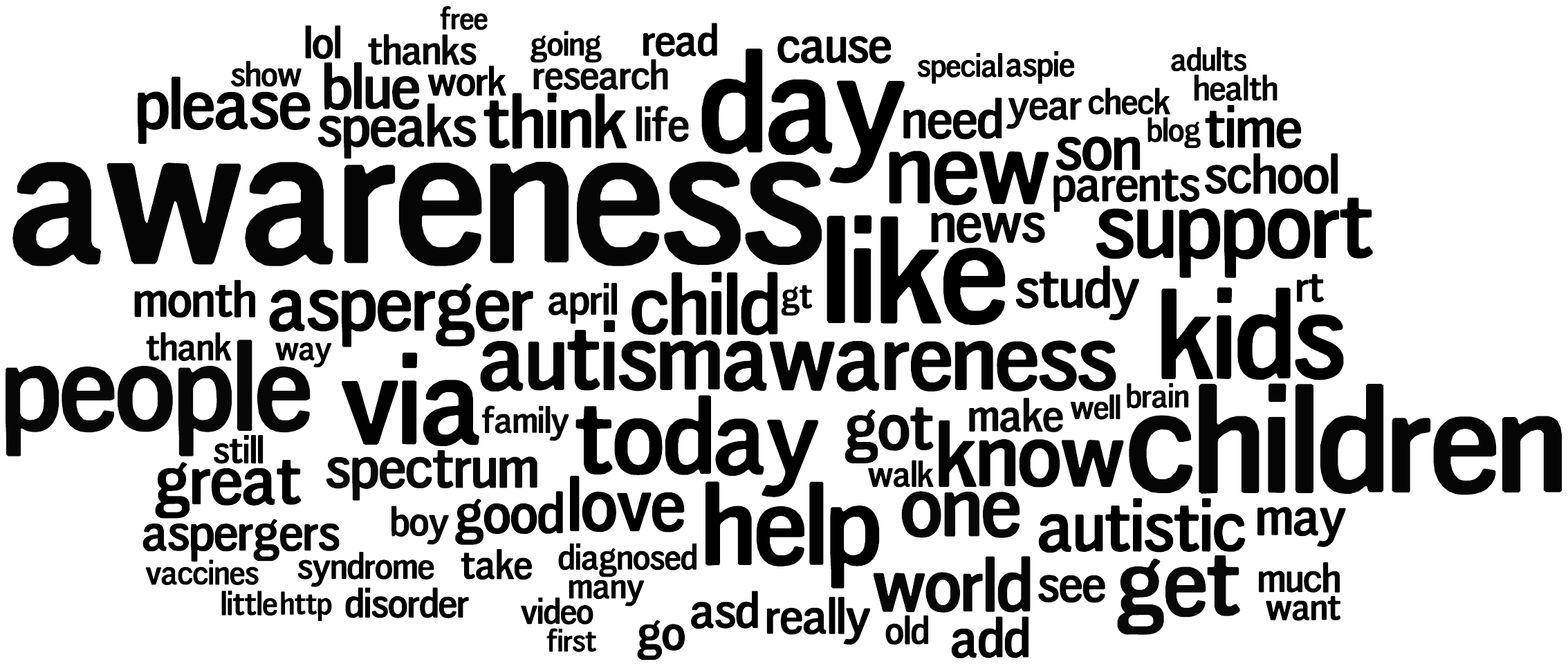}}
  \subfigure[Control `word-cloud']{\includegraphics[bb=520bp 250bp 2160bp 3000bp,clip,angle=-90,width=0.8\textwidth]{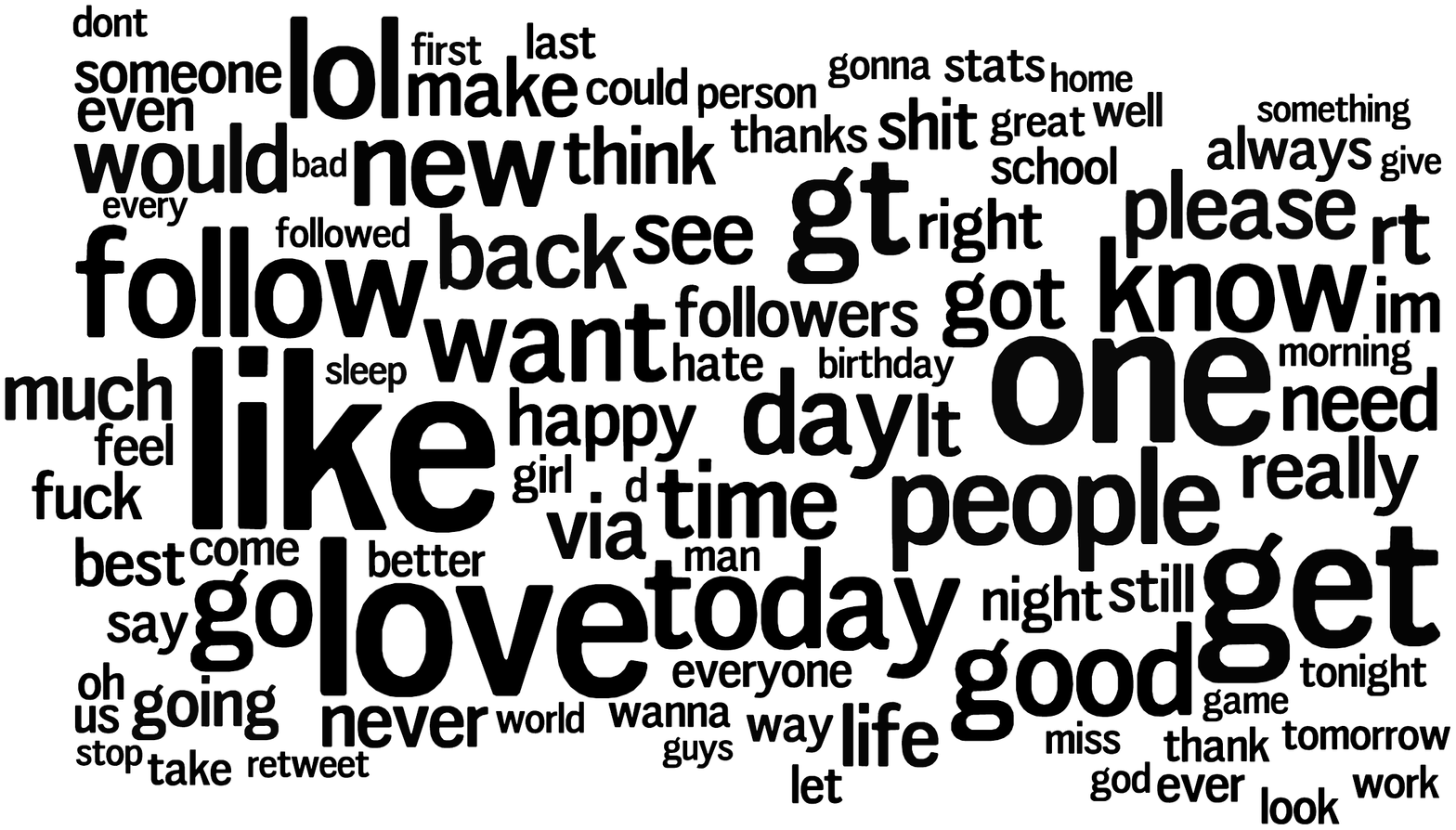}}
  \caption{ The most frequent words in the (a) ASD and (b) control data sets, shown as so-called `word-clouds'. The font size used to display a particular word is proportional to the corresponding word's frequency in the data set.}
  \label{f:TopWords}
\end{figure*}

In the first experiment we approached this task by comparing the most frequently used words in the two data sets. These words can be seen as a simple cumulative proxy for the actual content of individual messages. The key results are illustrated in Fig.~\ref{f:TopWords}(a) and Fig.~\ref{f:TopWords}(b) which show the 100 most frequent words in respectively the ASD and control data sets, displayed as so-called `word-clouds' whereby the frequency of a particular word is encoded by the corresponding font size. We used a linear scale, the font size thus being proportional to the corresponding word's frequency in a data set.

As readily observed from Fig.~\ref{f:TopWords} the most commonly used words in the two groups of tweets reveal a substantial difference in the nature of discussed topics. A more thorough examination reveals even further meaningful patterns coherent with the existing literature on ASD. In particular, observe the presence of a large number of words in the plot corresponding to the ASD group which are related to children such as ``children'', ``kids'', ``child'', and ``son'', for example. There are a number of reasons why this is unsurprising. Firstly, the diagnosis of ASD is usually made in early childhood so it is reasonable to expect that the parents confronting this new challenge would have increased initiative at seeking help from the community of parents in a similar situation. In addition, while undoubtedly always vulnerable, the vulnerability of individuals on the autism spectrum is the greatest while they are young which is when they need the most support from their guardians and therapists e.g.\ at acquiring the skills needed to progress thorough the educational system and integrate in the society.

It is also interesting to observe the high frequency of the words ``son'' and ``boy'' in the ASD data (top-right section of the word cloud), and the absence of the equivalent female sex words ``daughter'' and ``girl''. This example illustrates well how the content of tweets can be used to extract some rather subtle information. In particular in this case our findings are consistent with the understanding of the medical community and a body of evidence which shows that boys are nearly five times more likely to suffer from autism than girls~\cite{Fomb2009}.

In contrast, the most frequently used words in the control group do not seem to follow any particular pattern or focus of interest, and instead pertain to more general everyday interests and activities. Lastly, it is interesting to observe that the set of words in Fig.~\ref{f:TopWords}(a) appears to contain more nouns than Fig.~\ref{f:TopWords}(b), and fewer verbs. This suggests a different nature of information exchange in the ASD group which appears more focused on issues (syndromes, disorders, interventions, support, help, and so on) and individuals (mostly children), both being described by nouns, while the users who posted tweets of the control group seem to be more interested in what it is that they are or will be doing (like, want, go, think, and so on). We will explore this quantitatively in more detail in Section~\ref{sss:POS}.

\subsubsection{Hashtag analysis}\label{sss:Hash}
An alternative proxy for tweet content can be found in user-designated tags, the so-called `hashtags'. These can be recognized by the leading special character `\#' (the hash sign) and are by convention understood to be meta-data labels in some manner connected with the content of the tweet which contains them. Recall from Section~\ref{ss:data} that 2,260,284 or approximately 25\% of the collected tweets contain hashtags.

\begin{figure*}[htb]
  \centering
  \subfigure[ASD hashtags `word-cloud']{\includegraphics[bb=110bp 70bp 520bp 700bp,clip,angle=-90,width=0.8\textwidth]{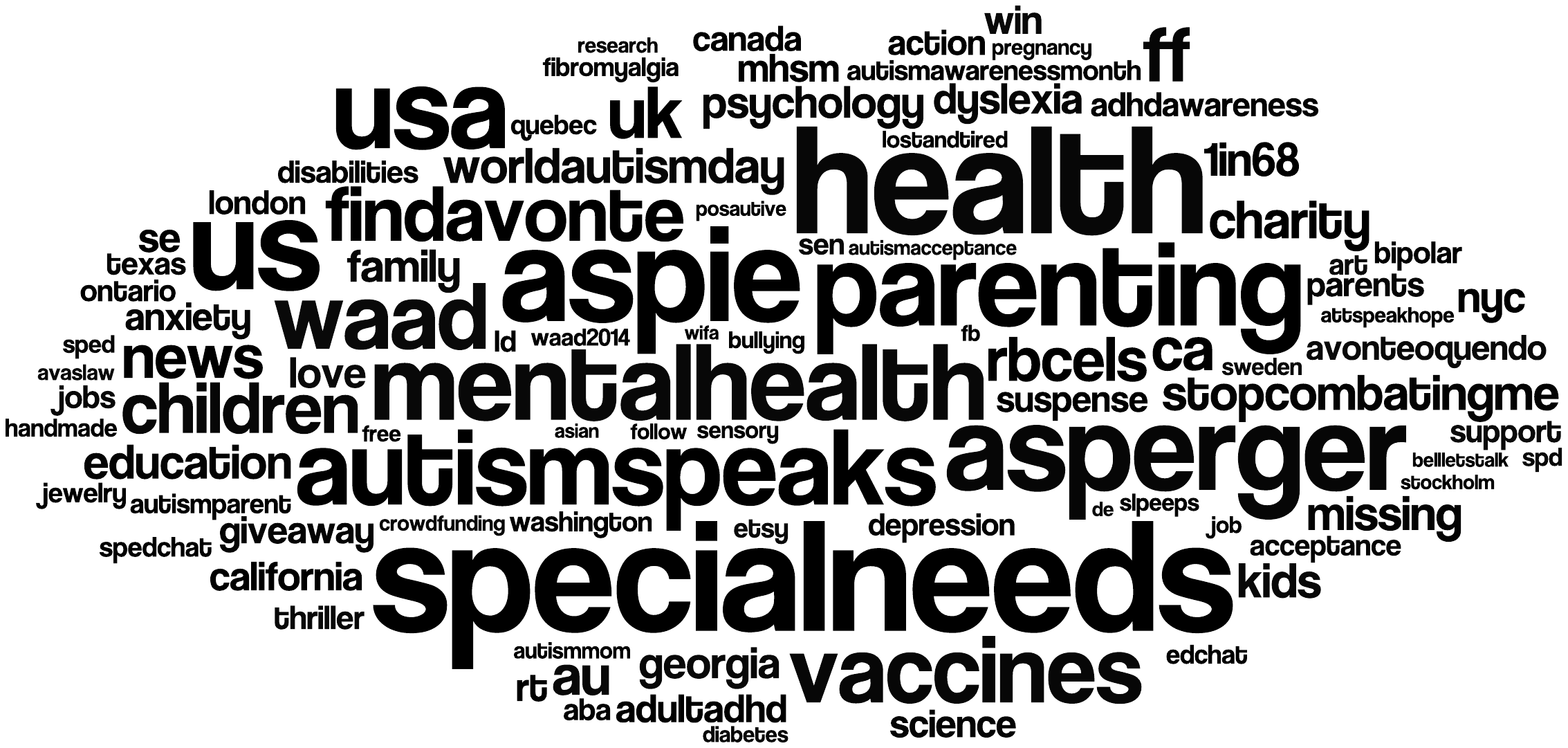}}
  \subfigure[Control hashtags `word-cloud']{\includegraphics[bb=120bp 70bp 500bp 700bp,clip,angle=-90,width=0.8\textwidth]{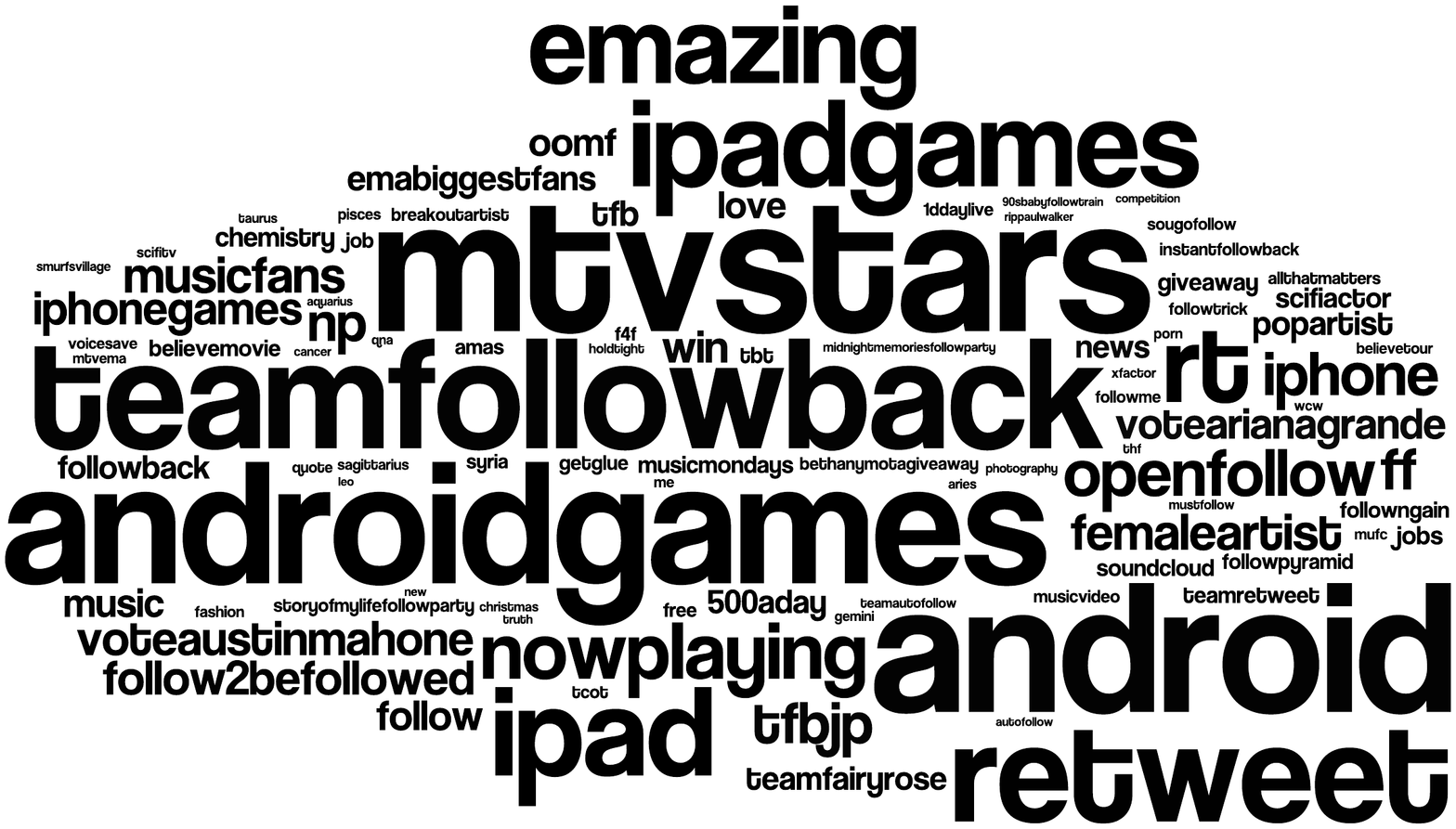}}
  \caption{The most frequent hashtags in the (a) ASD and (b) control data sets, shown as so-called `word-clouds'. The font size used to display a particular word is proportional to the corresponding word's frequency in the data set. Note that the hashtags corresponding to the search keywords have been removed before producing the ASD word-cloud. }
  \label{f:hashtags}
\end{figure*}

We summarize our findings as word-clouds of the 100 most frequent hashtags in the ASD and control data sets in Fig.~\ref{f:hashtags}(a) and Fig.~\ref{f:hashtags}(b). As in the previous analysis of the most frequently used terms, the difference between the two data sets is readily apparent. Unsurprisingly, most of the hashtags in the ASD data set pertain to health issues, some of which are \textit{ipso facto} ASD-related, such as ``mentalhealth'' and ``psychology'', while others are less obviously so. Examples of the latter include ``fibromyalgia'' and ``vaccines''. Fibromyalgia is a class of disorders related to the body's processing of pain which recent evidence suggests may have a potential connection with ASD~\cite{GeieKernDaviKing+2011}. Similarly, although now discredited, previous research had suggested a causal link between children being vaccinated and developing autism~\cite{WakeMurcAnthLinn+1998}. These examples are further evidence of the type of powerful information which can be harvested from Twitter. In particular, it shows that it is possible to data-mine tweets to provide feedback to medical practitioners on the concerns of the ASD community, the penetration (or lack thereof) of relevant public health recommendations (e.g.\ ABA or the applied behaviour analysis), or the adoption of treatments of questionable efficacy (e.g.\ homeopathy, gluten-free diet).

Much like the word-cloud in Fig.~\ref{f:TopWords}(a), the hashtag word-cloud in Fig.~\ref{f:hashtags}(a) contains many references to various topics pertaining to the support of individuals on the autism spectrum. However, the nature of the two sets of terms is somewhat different. While the most frequently used terms mostly referred to the issues themselves, the corresponding hashtags mostly refer to particular support groups. For example, ``autismspeaks'' (centre-left position within the word-cloud) is a US-based autism advocacy organisation, ``thisisautism'' (centre-right position within the word-cloud) a Twitter stream used to share autism-related experienced, and `AS2DC' (top-right corner of the word-cloud) an action summit on autism held in Washington D.\ C.

\subsubsection{Part-of-speech analysis}\label{sss:POS}
In Section~\ref{sss:WordFreq}, in the discussion of our findings in the comparative analysis of the most frequently used words in the ASD and control tweets, we observed that the ASD tweets appeared to contain a greater proportion of nouns and a smaller proportion of verbs than the control tweets. This observation led us to investigate this matter in further detail. In particular, we compared the usage of different part-of-speech (POS) types in the two data sets.

\begin{figure*}
  \centering
  \subfigure[Nouns]{\includegraphics[width=0.49\textwidth]{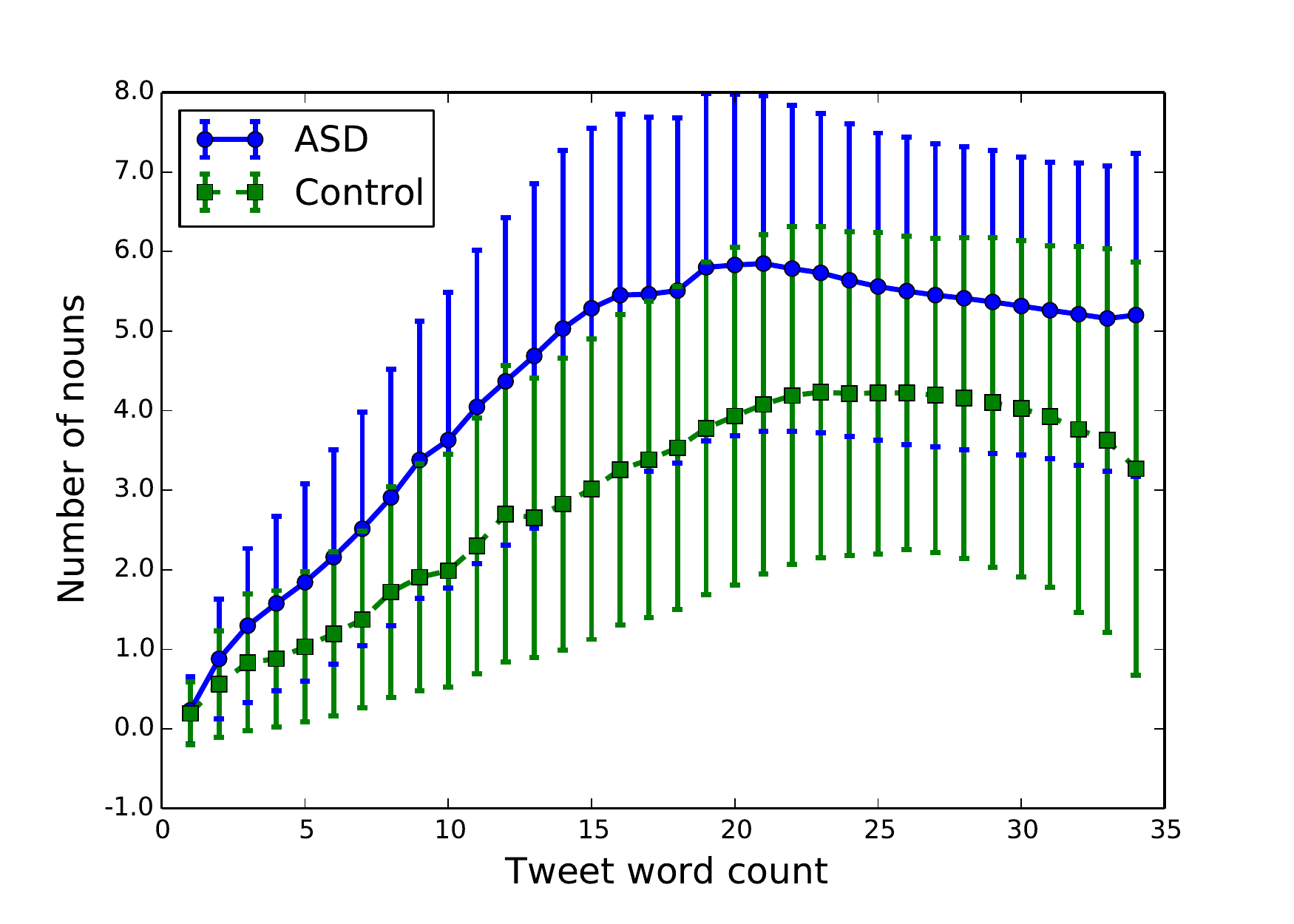}}
  \subfigure[Verbs]{\includegraphics[width=0.49\textwidth]{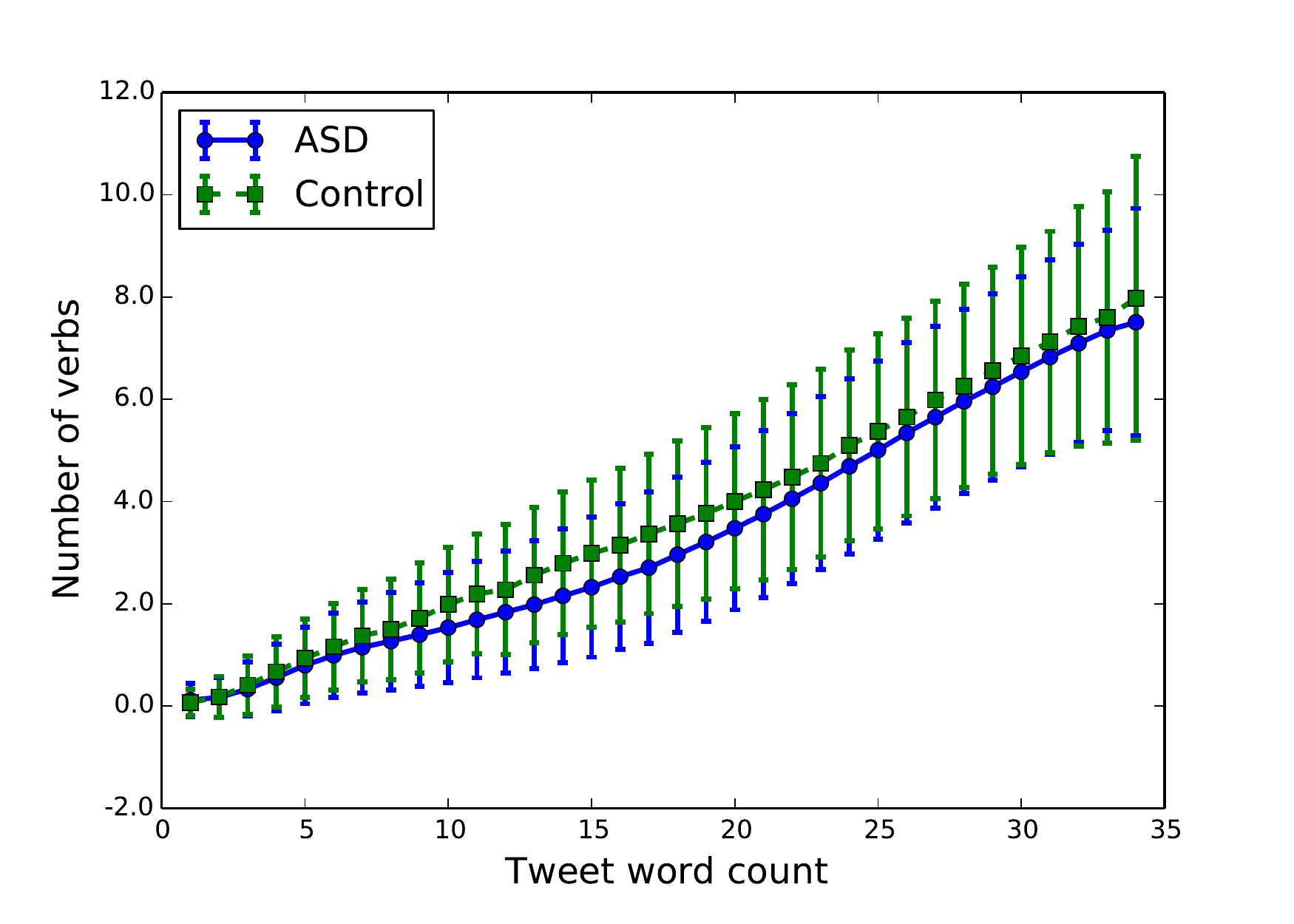}}
  \subfigure[Proper nouns]{\includegraphics[width=0.49\textwidth]{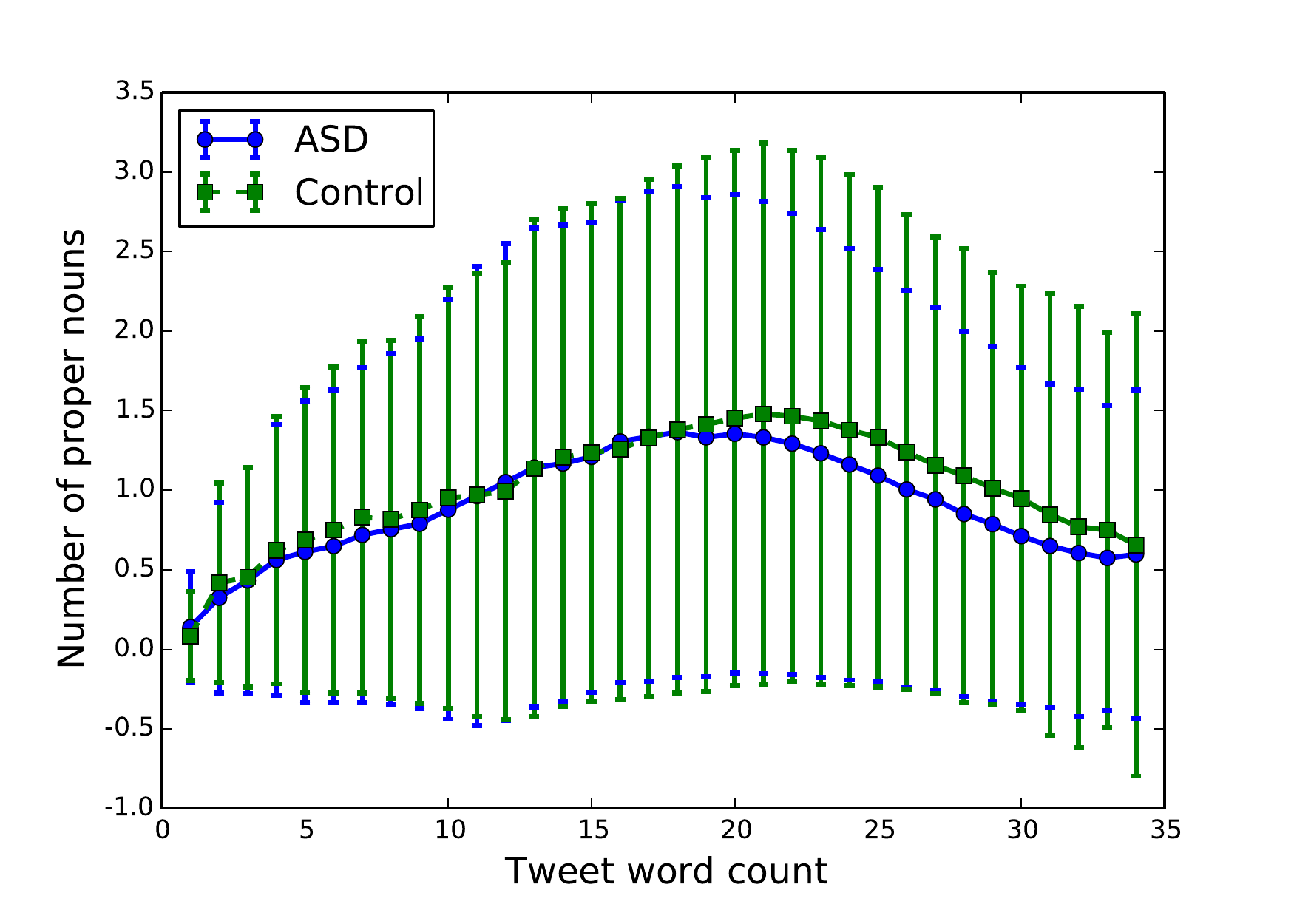}}
  \subfigure[Pronouns]{\includegraphics[width=0.49\textwidth]{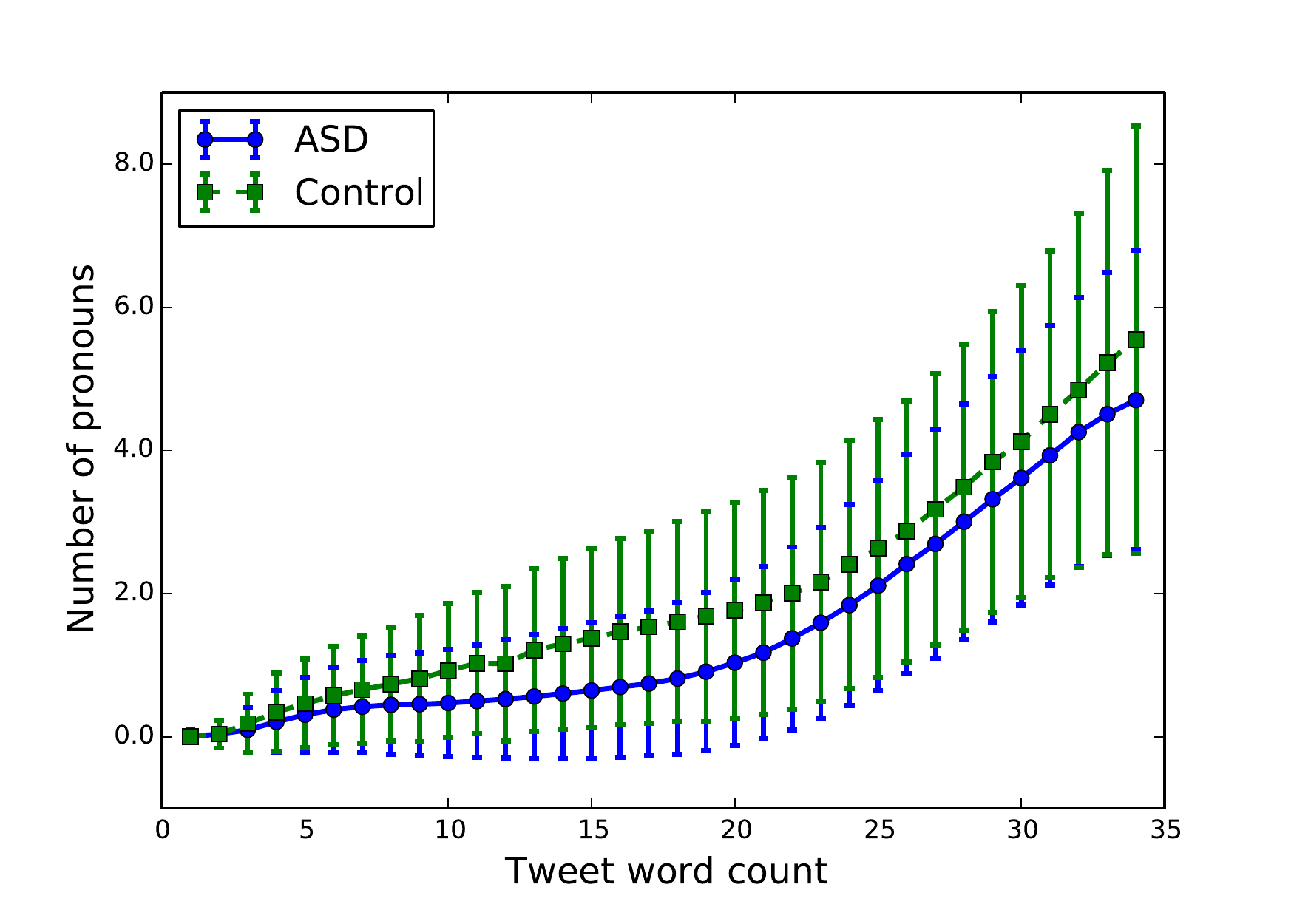}}
  \subfigure[Adjectives]{\includegraphics[width=0.49\textwidth]{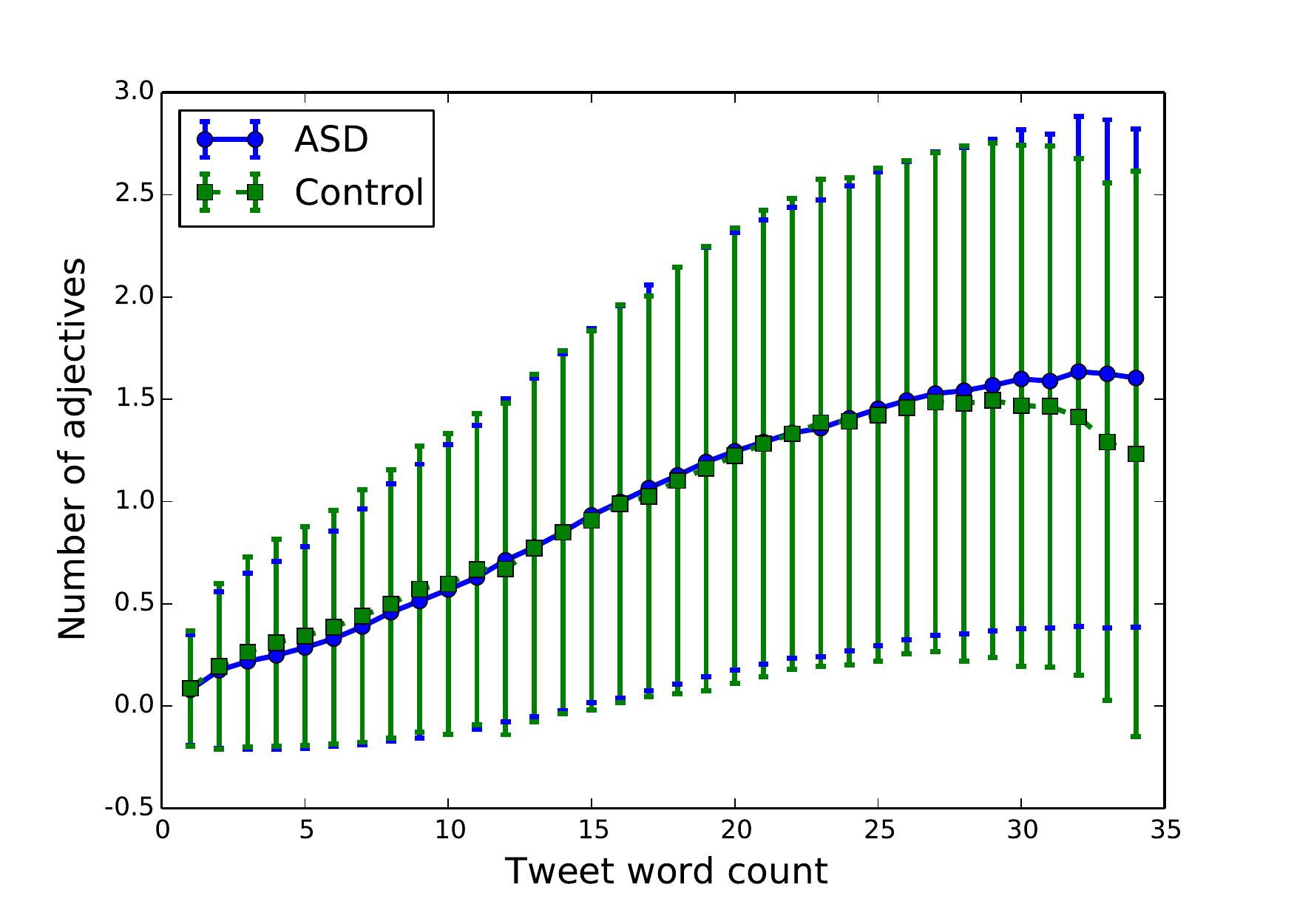}}
  \subfigure[Adverbs]{\includegraphics[width=0.49\textwidth]{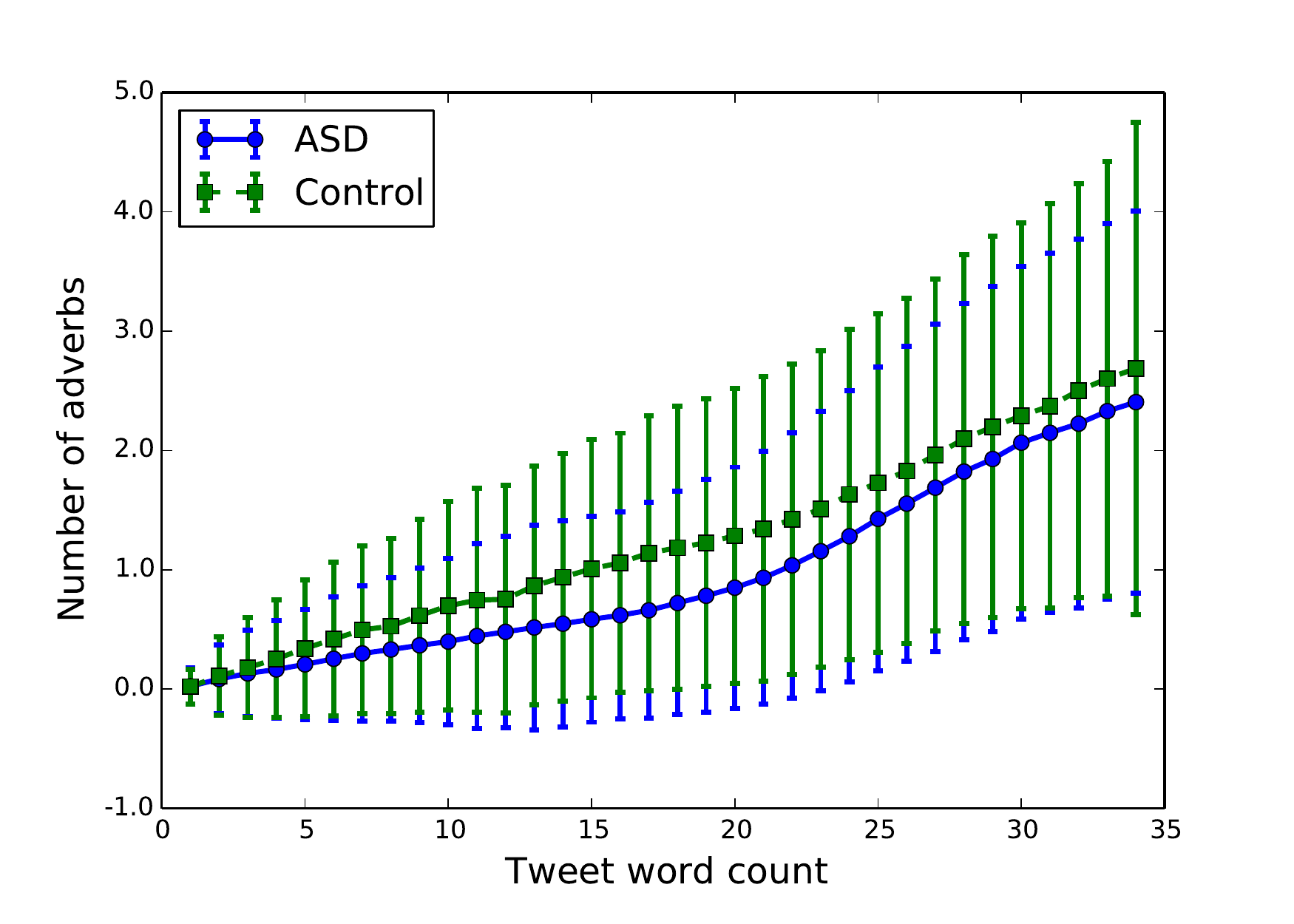}}
  \subfigure[Interjections]{\includegraphics[width=0.49\textwidth]{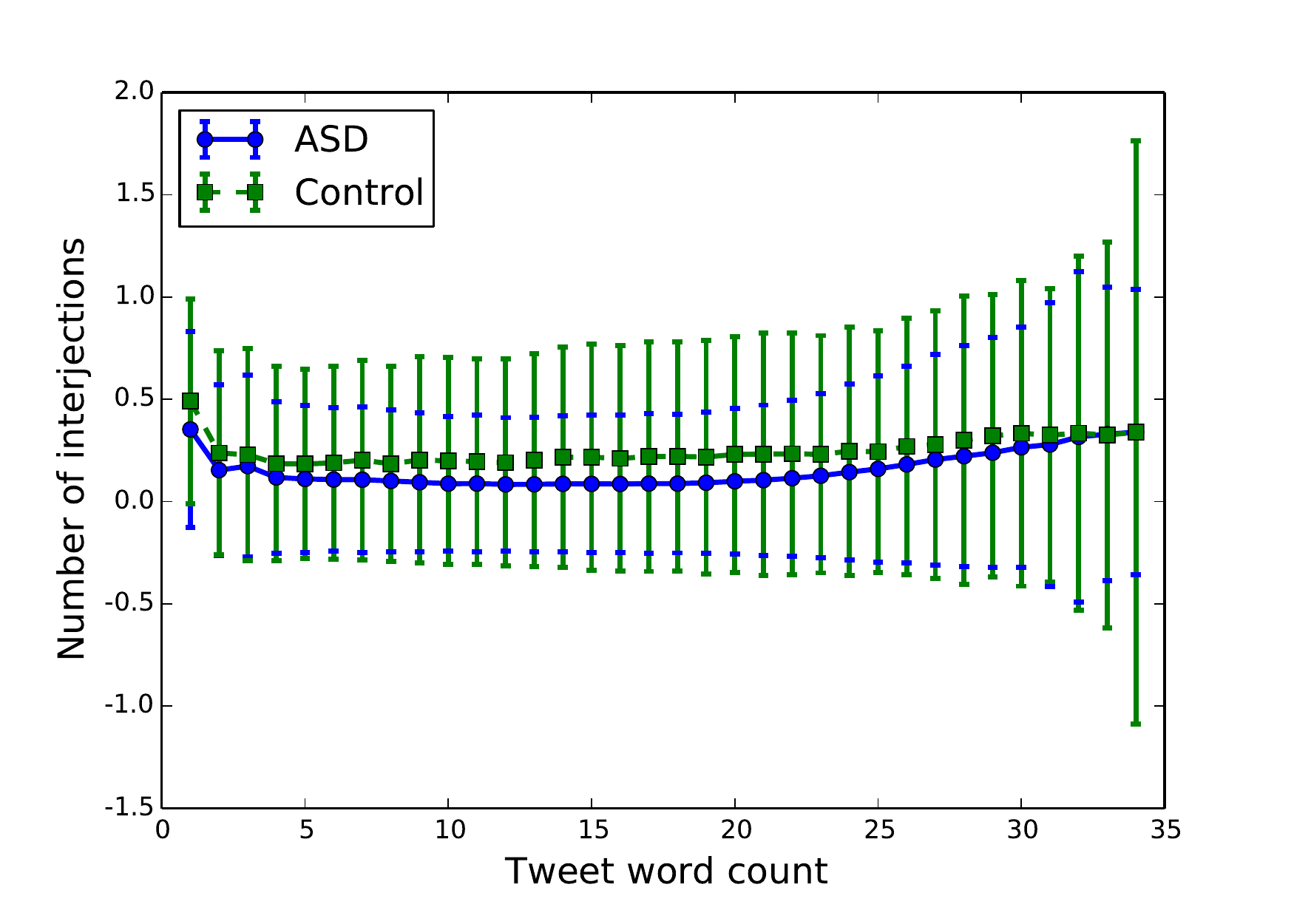}}
  \caption{A summary of the key results of comparative part-of-speech analysis. In all cases a statistically significant ($p<0.05$) difference between ASD and control sub-groups is observed. The most interesting insight is that tweets in the ASD sub-group are more focused on issues (syndromes, disorders, interventions, support, help, and so on) and individuals (mostly children), described by nouns, while the control tweets more often relate to what it is that their authors are doing (liking, wanting, going, thinking, etc). }
  \label{f:POS}
\end{figure*}

We used the free TweetNLP software package~\cite{OwopOConDyerGimp+2013}, trained on Twitter data, to tag automatically all terms in our entire data set according to their POS type. A selected set of the most interesting results is shown in the plots in Fig.~\ref{f:POS}. The first thing to observe from the entire corpus of results is that there is a consistent difference between the ASD and control tweets. Notwithstanding the large standard deviation values and the significant overlap of distributions corresponding to ASD and control tweets, that the difference between the two corpora is real rather than a result of stochasticity, is readily witnessed by the fact that in each of the plots the same relative behaviour is exhibited regardless of the tweet length. We confirmed this statistical significance rigorously too. In order to collate all data shown in a single plot, rather than considering the actual POS type count per tweet we considered POS type count normalized by tweet length. This allowed us to perform a single Student's $t$-test per POS type (i.e.\ plot in Fig.~\ref{f:POS}). In all cases we obtained $p<0.05$.

Next we turned our attention to the interpretation of the specific results in Fig.~\ref{f:POS}. Note by observing the plots in Fig.~\ref{f:POS}(a) and Fig.~\ref{f:POS}(b) that our hypothesis
from Section~\ref{sss:WordFreq} regarding the type of information communicated within the ASD and control sub-groups, is confirmed. Recall that by examining qualitatively word clouds of the dominant terms in the two corpora we hypothesised that the tweets in the ASD sub-group are more focused on issues (syndromes, disorders, interventions, support, help, and so on) and individuals (mostly children), described by nouns, while the control tweets more often relate to what it is that their authors are doing (liking, wanting, going, thinking, etc). Indeed for the average tweet length (as counted by the number of words in a message) of 15-20 terms, the number of nouns in the ASD data set is about twice the number of nouns in the control data set. The difference in the number of verbs, while not as substantial, is also significant -- for the average length message, the control data set contains approximately 20\% more verbs. Interestingly, the number of proper nouns does not appear to differ significantly between the two groups. This is perhaps somewhat surprising, as it could be expected that the control group would engage more in the discussion of so-called celebrities. A more thorough, manual examination of relevant tweets reveals the reason -- while the ASD group does engage in less chatter about celebrities, this is offset by the substantial attention that autistic people, specific autism activists, or politicians commenting or acting on autism-related issues receive.

\subsection{Tweet classification}\label{ss:Class}
In Section~\ref{sss:Zipf} we provided evidence that tweets, including those made by the community interested in the ASD, obey some of the same general regularities when it comes to the pattern of word usage. Then, starting in Section~\ref{ss:Length} and corroborating this further in Sections~\ref{sss:WordFreq},~\ref{sss:Hash}, and \ref{sss:POS}, we showed that in terms of their semantics and content, messages of the ASD community appear to exhibit a range of characteristics which differentiate them from those in our control data set. This motivated us to explore if it is possible to classify tweets automatically as belonging to the ASD data set or not.

It is important to emphasise right at the beginning that in this experiment we eliminated from all tweets the five search keywords which we used to divide our entire tweet data corpus into ASD and control sets i.e.\ which we used to \emph{define} the quasi-ground truth labelling. Recall from Section~\ref{ss:data} that these keywords are ``autism'', ``adhd'', ``asperger'', and ``aspie''. Had they not been removed, classification performance would have been artificially increased as their presence in a tweet would have been learnt as being capable of perfectly predicting classification output. Furthermore, it is reasonable to expect that the `true' ASD corpus of tweets should include some tweets which do not contain any of the aforementioned five keywords. In that sense, we hypothesised that it would be possible to use simple tweet filtering based on a small number of obvious keywords to build a bootstrap training data set which could be in turn used to learn additional aspects characterizing ASD-related tweets, thereby facilitatating a more robust retrieval of messages of interest.

\begin{figure}[htb]
  \centering
  \includegraphics[width=0.9\textwidth]{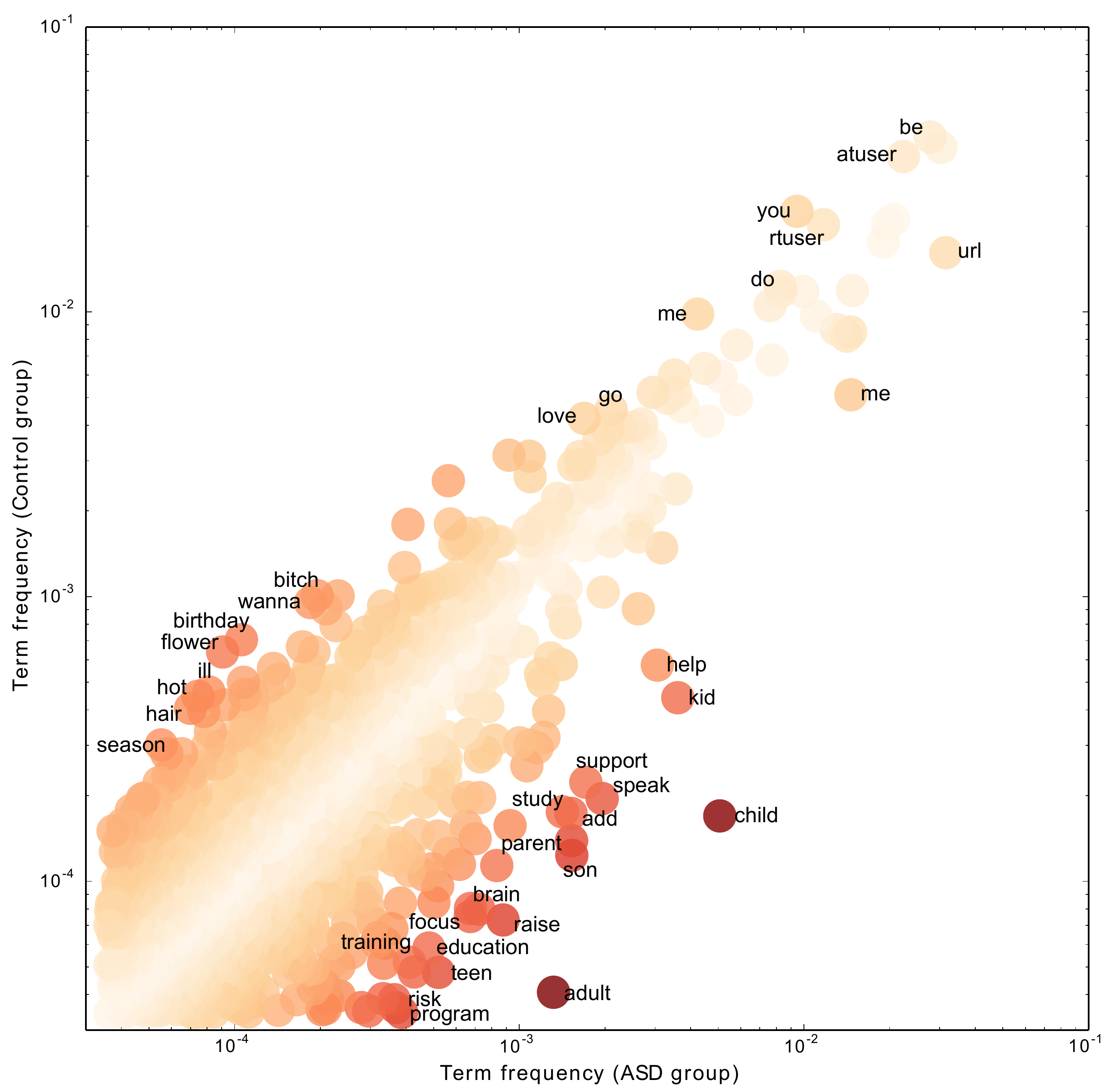}
  \caption{A compact illustration of the discriminative power of different individual words. Each circular blob represents a word, the saturation of its colour being proportional to the absolute value of the logarithm of the ASD-control likelihood ratio of the word. }
  \label{f:DiscriminativeWords}
\end{figure}

\paragraph{Discriminative power of individual terms}
To examine the discriminative potential of tweet content following the removal of our keywords, we first looked at the discriminative power of individual words. This is effectively visualized in Fig.~\ref{f:DiscriminativeWords}. What the plot shows is the most frequent 1300 words in our entire data set, which covers respectively 75\% and 83\% of the words in ASD and control subsets, depicted as circular blobs and colour-coded by their discriminative power (specifically, the saturation of a blob's colour is proportional to the absolute value of the logarithm of the ASD-control likelihood ratio of the corresponding word). As expected based on our previous results presented in Sections~\ref{sss:WordFreq} and~\ref{sss:Hash}, a number of words pertaining to issues of most concern to the ASD community are highly discriminative. This observation motivated our next set of experiments which examined tweet classification in detail.

\subsubsection{Automatic classification}

\paragraph{Representations}
We evaluated a number of different representations of tweet content. Here we present three which overall produced the most interesting results. These are variations on well-known representations in the existing literature, adapted to the problem at hand. Specifically, we report on the performance of the following:
\begin{itemize}
  \item binary bag of terms (pre-processed words)~\cite{LewiRing1994,Aran2015c},\\[-5pt]
  \item integer bag of terms (also referred to as term count)~\cite{Aran2010}, and\\[-8pt]
  \item tf-idf (term frequency-inverse document frequency) score~\cite{Robe2004}.\\
\end{itemize}

In the binary bag of terms representation used in this work, each entry $x_i$ in a feature vector $\mathbf{x}_\text{bow}=\left[x_1,x_2,\ldots,x_{n_w} \right]^T$ corresponds to a particular term and is coded as either present in a particular tweet (value 1) or absent from it (value 0), where $\mathbf{x}_\text{bow} \in \mathbb{R}^{n_d}$ and $n_d$ is the dictionary size over which the feature vector is constructed (we will discuss vocabulary construction in more detail shortly). Thus, the original tweet:
\begin{quote}\it
  Looks like we will have more \#autism research happening for children in \#EarlyIntervention next year! :-) \#VisualSupports \#MobileTech
\end{quote}
which following our pre-processing ends up as:
\begin{quote}\it
  look like autism research happen child in earlyintervention visualsupports mobiletech
\end{quote}
is represented using the following binary bag of words feature vector:
\begin{align}
  \mathbf{x}_\text{bow}=\bigg[~0 ~~ \ldots ~~ \overbrace{1}^{\text{look}} ~~ 0 ~~ \ldots ~~ \overbrace{1}^{\text{like}} ~~ 0 \ldots ~~ \overbrace{1}^{\text{research}} ~~ 0 ~~ \ldots ~~ 0 ~~ \overbrace{1}^{\text{happen}} ~~ 0 \ldots ~~ \overbrace{1}^{\text{child}} ~~ 0 ~~ \ldots ~~ 0~\bigg]^T \notag
\end{align}
Note that the term ``autism'' is not included as it is matched by one of our search keywords used for quasi-ground truth labelling, as described in Section~\ref{ss:data}. On the other hand, the terms ``earlyintervention'', ``visualsupports'', and ``mobiletech'' do not have the corresponding entries in the feature vector because their frequency across the data corpus is too low i.e.\ they are not amongst the most frequent terms included in our vocabulary. Lastly, ``in'' is excluded (automatically, as all others) on the basis that it is too short.

The term count representation is similar to the binary bag of words described previously. As before each entry in a feature vector corresponds to a particular term. However, unlike before the entries are not binary variables which signify the presence or the absence of a particular term but rather actual counts of the term's instances in a tweet. Since the previously given example of a tweet does not include any repeated terms, the term count feature vector $\mathbf{x}_\text{tc} \in \mathbb{R}^{n_d}$ in this instance is identical to $\mathbf{x}_\text{bow}$. On the other hand, the tweet mentioned in Section~\ref{sss:Preprocess} which after pre-processing looks as follows:
\begin{quote}\it
  101 autism genet analysi individu autism find gene delet use power genet sequenc  url
\end{quote}
produces different bag of words and term count feature vectors, the former having the value of 1 for the entry corresponding to the term ``genet'' and the latter the value of 2 since the term has two occurrences in the tweet.

Lastly, the tf-idf representation $\mathbf{x}_\text{tfidf} \in \mathbb{R}^{n_d}$ too has entries which correspond to different vocabulary terms, but with values which measure the importance of a particular term in a tweet. Specifically, a particular entry is equal to the frequency of the corresponding term (`term frequency') weighted by the inverse number of tweets in the training corpus which also contain the term (`document frequency'). This representation has been used successfully in a variety of applications, from text mining~\cite{Robe2004} to visual object recognition and retrieval~\cite{Aran2012f}.

\paragraph{Vocabulary construction}
We also explored a number of different ways of selecting the vocabulary over which tweet feature vectors are constructed based on the discriminativeness of terms, as well as their group-based or combined frequencies of use. Specifically, we examined vocabularies formed by choosing the most frequent terms in one of the two sub-groups, the most discriminative terms in one of the two sub-groups (as explained in Section~\ref{ss:Class} and visualised in Fig.~\ref{f:DiscriminativeWords}), the union or the intersection of either of the former (e.g.\ the union of the most discriminative terms in the ASD sub-group and the most discriminative words in the control sub-group), the most frequent words across the entire corpus (i.e.\ both sub-groups), and the most discriminative words across the entire corpus. Our results suggest that the best performance is achieved by selecting a set of the overall most frequent terms; due to the limitations of space we report the corresponding results only, using the 1500 most frequent terms.

\paragraph{Classification methodology}
In all experiments reported in this section we adopt the supervised classification paradigm. Specifically, we assume that we have available a training set of pairs $\{(\mathbf{x}_1^{(t)},y_1^{(t)}),(\mathbf{x}_2^{(t)},y_2^{(t)}),\ldots,(\mathbf{x}_{n_t}^{(t)},y_{n_t}^{(t)})\}$ where $\mathbf{x}_i^{(t)}$ is the feature vector (representation) corresponding to the $i$-th training tweet and $y_i^{(t)}$ its binary label signifying if the tweet belongs to the ASD or the control sub-group. After a classifier is trained the label of a novel query tweet described by the feature vector $\mathbf{x}$ is performed as follows:
\begin{align}
  y = \arg \max_{y} \frac{Pr(y)Pr(\mathbf{x}|y)}{Pr(\mathbf{x})} = \arg \max_{y} Pr(y)Pr(\mathbf{x}|y).
\end{align}
In our experiments we used half of the collected data corpus for training, and the remaining half for testing the performance of different representations and classifiers. Three popular classification methods were examined:
\begin{itemize}
  \item na\"{\i}ve Bayes-based~\cite{Bish2007},\\[-5pt]
  \item logistic regression-based~\cite{YuHuanLin2011}, and\\[-8pt]
  \item least absolute shrinkage and selection operator (LASSO)-based.\\
\end{itemize}

In na\"{\i}ve Bayes classification the strong assumption of independence between different terms in a feature vector is assumed resulting in the following class likelihood estimate:
\begin{align}
  P_{NB}(y=\pm 1|\mathbf{x}) \propto \prod_{i=1}^{n_w} Pr(x_i | y=\pm 1),
\end{align}
where, without loss of generality, the values +1 and -1 of the dependent variable $y$ are used to signify respectively ASD and control sub-group memberships. Conditional probabilities $Pr(x_i | y=\pm 1)$ are estimated in a straightforward manner from the frequencies of the corresponding terms in the training data corpus.

In logistic regression, the conditional probability of the dependent variable $y$ is modelled as a logit-transformed multiple linear regression of the explanatory variables $x_1,\ldots,x_{n_e}$:
\begin{align}
  P_{LR}(y=\pm1|\mathbf{x},\mathbf{w})=\frac{1}{1+e^{-y\mathbf{w}^T\mathbf{x}}}.
\end{align}
The model is trained (i.e.\ the weight parameter $\mathbf{w}$ learnt) by maximizing the likelihood of the model on the training data set, given by:
\begin{align}
  \prod_{i=1}^{n_t} Pr(y_i^{(t)}|\mathbf{x}_i^{(t)},\mathbf{w}) = \prod_{i=1}^{n_t}\frac{1}{ 1+e^{-y_i^{(t)}\mathbf{w}^T\mathbf{x}_i^{(t)}} },
\end{align}
penalized by the complexity of the model:
\begin{align}
  \frac{1}{\sigma\sqrt{2\pi}}e^{-\frac{1}{2\sigma^2}\mathbf{w}^T \mathbf{w}},
\end{align}
which can be restated as the minimisation of the following regularized negative log-likelihood:
\begin{align}
  \mathfrak{L}=C \sum_{i=1}^{n_t}\log\left( 1+e^{-y_i\mathbf{w}^T\mathbf{x}_i} \right)+\mathbf{w}^T\mathbf{w}.
  \label{e:loglLR}
\end{align}
A coordinate descent approach described by Yu \textit{et al.}~\cite{YuHuanLin2011} was used to minimize $\mathfrak{L}$.

Finally, in LASSO-based classification, sparseness of the solution is achieved by replacing the $L_2$-norm penalty in \eqref{e:loglLR} with $L_1$-norm, yielding the following regularized negative log-likelihood:
\begin{align}
  \mathfrak{L}=C \sum_{i=1}^{n_t}\log\left( 1+e^{-y_i\mathbf{w}^T\mathbf{x}_i} \right)+\lambda \|\mathbf{w}\|_1.
\end{align}
where $\lambda$ is a free parameter governing the tradeoff between prediction loss and solution sparsity.

\paragraph{Basic classification: results and discussion}
We first evaluated the na\"{\i}ve Bayes and logistic regression-based classifiers as these do not have free parameters i.e.\ parameters which must be set \textit{a priori}. The corresponding classification results are summarized in Table~\ref{t:res}. With the exception of the result achieved with logistic regression and the tf-idf representation, different combinations of base classifiers and representations performed nearly identically. Specifically, it can be seen that approximately 79\% of tweets were correctly classified. This is a rather remarkable performance considering the brevity of tweets and the fact that some of the most informative (in the context of the classification task at hand) words were not used for classification. It is insightful to notice the consistently higher accuracy attained for control (84-85\%) rather than ASD tweets (71-74\%). We explored this finding in more depth by manually inspecting misclassified messages. Within the ASD data corpus, we found that the main source of classification error lied in the absence of any ADS-specific information in very short tweets, after the removal of our four keywords used to construct the data set in the first place (e.g.\ ``it is the adhd, oops!''). This is a highly comforting finding since of course in any practical application these keywords would not be eliminated, thus improving classification performance dramatically.

\begin{table*}[htb]
  \centering
  \renewcommand{\arraystretch}{1.5}
  \caption{A summary of the classification results (ASD-related vs.\ non-ASD-related tweets). Each cell (corresponding to a combination of a classification method and a representation) shows the associated confusion matrix. The first row/column of a confusion matrix corresponds to the control class and second row/column to the ASD class (thus for e.g.\ using the term count representation, na\"{\i}ve Bayes correctly correctly classified 84\% of control tweets and 71\% of ASD tweets).}
  \vspace{10pt}
  \begin{tabular}{|l|c|c|c|}
    \cline{2-4}
    \multicolumn{1}{c|}{} & \multicolumn{3}{c|}{Confusion matrix}\\
    \hline
    Representation & Binary & Term count & tf-idf\\
    \hline
    Na\"{\i}ve Bayes & $\begin{array}{cc}
                  0.84 & 0.16\\
                  0.29 & 0.71
                \end{array}$
             & $\begin{array}{cc}
                  0.84 & 0.16\\
                  0.29 & 0.71
                \end{array}$
             & $\begin{array}{cc}
                  0.84 & 0.16\\
                  0.38 & 0.62
                \end{array}$\\
    \hline
    Logistic regression & $\begin{array}{cc}
                           0.85 & 0.15\\
                           0.26 & 0.74
                         \end{array}$
                      & $\begin{array}{cc}
                           0.85 & 0.15\\
                           0.27 & 0.73
                         \end{array}$
                      & $\begin{array}{cc}
                           0.03 & 0.97\\
                           0.04 & 0.96
                         \end{array}$\\
    \hline
  \end{tabular}
  \label{t:res}
\end{table*}

We also experimented with the application of latent semantic analysis (LSA)~\cite{Duma2004}, testing out the possibility that better discrimination may be achieved by working in the so-called concept rather than term space. Although LSA has been shown to be highly successful in a variety of information retrieval tasks, our experiments suggest that the same benefit is not observed when it is applied on tweets. The likely reason for this can be found in the brevity of tweets, which favours the use of succinct expressions whereby individual terms (unigrams) themselves effectively become LSA concepts. This also explains why the simple na\"{\i}ve Bayes classifier performed on par with the generally superior logistic regression-based one, as can be seen in Table~\ref{t:res}.

\begin{figure}[htbp]
  \centering
  \includegraphics[width=0.9\linewidth]{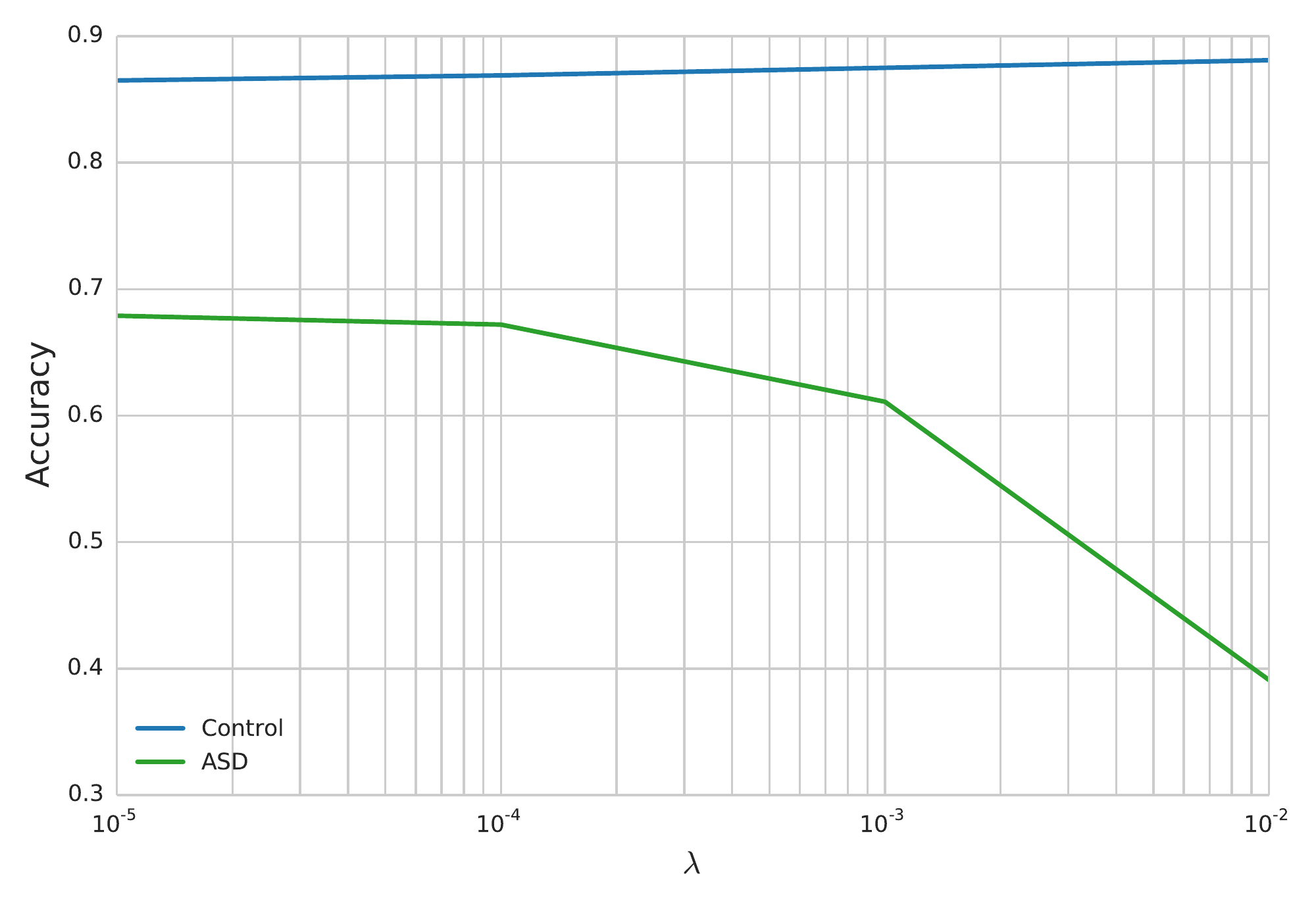}
  \caption{LASSO-based classification performance as a function of the loss-sparseness tradeoff parameter $\lambda$. The number of non-zero regression coefficients for $\lambda$ equal to $10e^{-5}$, $10e^{-4}$, $10e^{-3}$, and $10e^{-2}$ is respectively 801, 580, 151, and 12. High values of $\lambda$ which severely penalize model complexity can be seen to result in deteriorating classification performance for ASD-related tweets. }
  \label{f:lasso}
\end{figure}

We next evaluated the LASSO-based classifier. In order to examine fully the behaviour of this approach we did not perform the usual procedure for selecting the value of the free parameter $\lambda$ using cross-validation; rather we measured classification performance across a range of $\lambda$ values. Specifically we considered values from $\lambda=10e^{-5}$ up to $\lambda=10e^{-2}$; greater values than $10e^{-2}$ produced numerical problems caused by an excessively small set of selected predictor terms. Our results are summarized in Fig.~\ref{f:lasso}. The plot shows that the classifier exhibits consistent performance across the range of $\lambda$ between $10e^{-5}$ and $10e^{-4}$, with the accuracy of the classification of ASD tweets deteriorating thereafter (with a small increase in the accuracy of control tweet classification). This deterioration is readily explained by examining the number of terms selected by LASSO i.e.\ by the number of terms with non-zero corresponding regression coefficients. For $\lambda=10e^{-5}$ this number is 801, dropping down for $\lambda=10e^{-4}$ to 580, and reducing drastically to 151 for $\lambda=10e^{-3}$ and to 12 for $\lambda=10e^{-2}$ as a consequence of the harsh penalty on model complexity effected by the large values of $\lambda$.

Lastly, useful insight can be gained by considering the terms which correspond to the largest magnitude regression coefficients. These are, in order, ``help'', ``kid'', ``aw'' (the automatically constructed stem of words such as ``awareness'' and ``aware''), and ``child''. The corresponding regression coefficients are all positive i.e.\ the presence of these terms predicts ASD-related tweets. Consequently, control tweets are predominantly predicted by the absence rather than presence of specific terms; the highest magnitude negative coefficient, that is coefficient which corresponds to the term which predicts a control tweet, is ``follow''. The observation that control tweets are predicted by the absence of ASD-related terms explains why a reduction in model complexity (i.e.\ a reduction in the number of ASD-terms considered by the classifier) does not produce a deterioration in classification performance on control tweets (rather, a small improvement can be noticed) while a major negative effect is readily observed for the ASD corpus.
% reflecting the observation that

\paragraph{Condition-oriented classification: results and discussion}
To motivate our next set of experiments, consider a classifier which distinguishes between medical condition-related and other tweets. Let the classifier correctly classify a tweet of the former type (dependent variable value $+1$) with the probability $p_m$ and of the latter type with the probability $p_n$ (dependent variable value $-1$). Imagine applying this classifier on our data set but interpreting its output as classifying tweets as being ASD related ($+1$) or not ($-1$). Since all ASD tweets are medical condition related, the probability of the classifier output being $+1$ given an ASD tweet is also $p_m$. On the other hand, the control corpus contains both tweets not related to a medical condition as well as those which are related to medical conditions other than ASD. Hence, the probability of the classifier output being $-1$ given a tweet from our control corpus is:
\begin{align}
  p_n(1-\rho_m)+(1-p_m)\rho_m
  \label{e:medClass}
\end{align}
where $\rho_m$ is the proportion of control tweets which are medical condition-related. The first term, $p_n(1-\rho_m)$, is the probability of a control tweet not related to a medical condition being selected as input to the classifier (probability $1-\rho_m$) and correctly classified (probability $p_n$). The second term, $(1-p_m)\rho$, is the probability of a medical condition related control tweet being selected as input (probability $\rho_m$) and \emph{incorrectly} classified (probability $1-p_m$), as the classifier output of interest is $-1$. The expression in \eqref{e:medClass} can be rearranged to give:
\begin{align}
  p_n+(1-p_m-p_n)\rho_m
\end{align}
Since the proportion $\rho_m$ of medical condition related tweets is likely to be small, it can be seen that a good classifier which differentiates between medical condition related and other tweets can appear to be a good classifier of ASD vs.\ non-ASD-related tweets. Thus the question is if the results of our experiments described in the previous section really demonstrate good ASD vs.\ non-ASD discrimination, or is it simply the case that our methods learnt to distinguish between medical condition-related and non-medical condition-related tweets. We set out to investigate this next.

For this experiment we followed the methodology described in the previous set of experiments with the sole difference being that learning was done using a different control corpus. Specifically, we collected a data set related to another pervasive medical condition -- Alzheimer's disease. In a similar manner as in Section~\ref{ss:data} we collected tweets which contain the word ``alzheimer'' (or any of the words derived from it by suffixation), balancing their number with the number of ASD-related tweets in our data set. Considering the little difference in performance obtained by the use of different representations in our previous set of experiments, we did not examine them all here too; instead we adopted the term count representation as the most commonly used one in the literature.

Our results are summarised in Table~\ref{t:resAlz}. It can be readily observed, especially in the context of the evidence presented previously, that this experiment too strongly supports our thesis that the content of ASD tweets, even with the most salient words removed (the keywords used to collect the corpus as described in Section~\ref{ss:data}) is highly informative. Both Alzheimer's disease-related and ASD-related tweets were classified with nearly perfect accuracy. As in our previous experiments there was little difference between the performance of the na\"{\i}ve Bayes classifier and that based on logistic regression, the latter achieving somewhat better results on ASD-related tweets (100\% vs. 98\% accuracy).

\begin{table*}[htb]
  \centering
  \renewcommand{\arraystretch}{1.5}
  \caption{A summary of the classification results (ASD-related vs.\ Alzheimer disease-related tweets). Each cell shows the associated confusion matrix. The first row/column of a confusion matrix corresponds to the Alzheimer's disease class and second row/column to the ASD class. }
  \vspace{10pt}
  \begin{tabular}{|l|c|}
    \cline{2-2}
    \multicolumn{1}{c|}{} & Confusion matrix\\
    \hline
    Representation & Term count \\
    \hline
    Na\"{\i}ve Bayes & $\begin{array}{cc}
                  0.98 & 0.02\\
                  0.05 & 0.95
                \end{array}$\\
    \hline
    Logistic regression & $\begin{array}{cc}
                           0.98 & 0.02\\
                           0.00 & 1.00
                         \end{array}$\\
    \hline
  \end{tabular}
  \label{t:resAlz}
\end{table*}

\subsection{Bootstrap keyword set analysis}
Recall from Section~\ref{ss:data} that we collected the ASD tweet corpus by including in it any tweet which contained any of the four keywords ``autism'', ``adhd'', ``asperger'', and ``aspie'' (or any of their derivatives obtained by suffixation). Moreover, in all experiments in which the difference between tweets related and unrelated to ASD was examined, the ground truth was \emph{defined} by the presence of these words. This was necessitated by the scale of the data which made it practically impossible to perform labelling manually. Consequently, these keywords had to be excluded from the consideration in classification experiments. Clearly this would not be done in any practical application which means that the results we presented are worse, likely significantly so, than they would have been had we been able to use the entirety of tweet content. The high classification accuracy we obtained of approximately 80\% provides strong evidence that even with these most salient words removed, the remaining content of ASD-related tweets is highly informative and characteristic. In the final set of experiments we report here we examined the sensitivity of learning to the exact keywords used to select the learning corpus.

\begin{figure}[htb]
  \centering
  \includegraphics[width=0.7\textwidth]{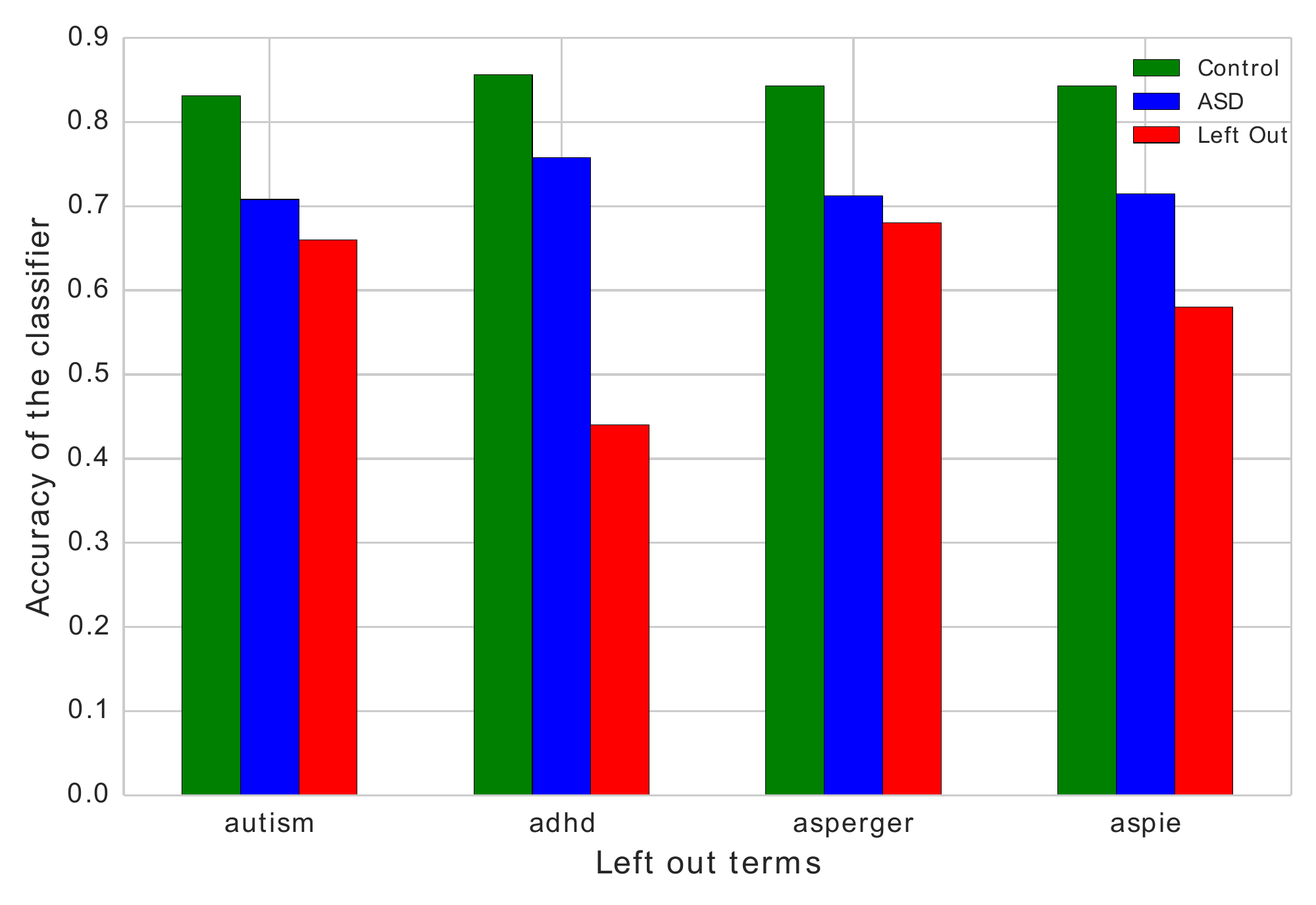}
  \caption{ Classification results using only a subset of the entire training corpus used in the previous section. In each experiment training was performed using tweets matched by the three remaining keywords after one of the original keywords was left out. For each experiment we show three classification accuracies: on (i) test corpus control tweets (green bars), (ii) test corpus ASD-related tweets (blue bars), and (iii) tweets matched by the left out keyword but no others (red). See main text for a discussion of the findings. }
  \label{f:leaveOut}
\end{figure}

We repeated the classification experiments from the previous section but using for training only a part of the previous training corpus -- the part that is matched by three of the four keywords. Thus four experiments were performed, with one of the keywords being left out in each. Then we first evaluated the performance of thus learnt classifier on the test corpus, just as before. We additionally evaluated the performance of the classifier specifically on those tweets which contained the left out keyword but none of the other three. Our results are summarized in the bar plot in Fig.~\ref{f:leaveOut}. Firstly, observe that in no experiment was the classification performance on the test set greatly negatively affected -- in all cases 83-85\% of control tweets were correctly classified and 71-76\% of ASD-related tweets. This is in line with our previous observations and is highly reassuring since it suggests that the same type of characteristic context is learnt regardless of which specific keywords are used to collect training data. However, the classification results on the tweets only matched by the left out keyword is interesting and provides novel insight. In particular, observe the particularly bad performance in the experiment in which the keyword ``adhd'' was left out -- only about 44\% of the tweets matched by this keyword were correctly classified. A manual examination of the misclassified tweets readily revealed the answer -- in nearly all cases these seemingly misclassified tweets indeed were not related to ASD. Rather the term ``adhd'' was used as a hyperbole for an inattentive or easily distracted person. Here are a few examples:
\begin{itemize}
\item \begin{quote}\it
  Watching football is such a commitment. I'm too adhd for this.
\end{quote}
\item \begin{quote}\it
  I think we both have adhd
\end{quote}
\item \begin{quote}\it
  I feel like I have adhd.. I cannot write 300 words and not stop for a ``break''
\end{quote}
\item \begin{quote}\it
  And then your adhd causes you to forget everything as soon as you get on your phone
\end{quote}
\end{itemize}
We made a similar observation with regard to the usage of the word ``aspie'' which too is used loosely in colloquial speech (albeit with a lower frequency than ``adhd''). This explains the performance of the classifier trained on data matched by the other three keywords and evaluated on the tweets matched by ``aspie'' only, which was also worse than the corresponding performance in the experiments in which the left out keyword was ``autism'' or ``asperger'' but not as much as when ``adhd'' was left out.

In summary, this experiment offers several important insights useful in the practical application of the ideas presented in the present paper. Firstly, it showed that the choice of keywords does not need to be a complex process in the sense that a small number of keywords is sufficient to learn the relevant salient characteristics of a theme associated with them. Additional keywords appear to be redundant, adding no new information and not resulting in improved performance. Secondly, the experiment demonstrated how potentially bad keyword choices can be automatically detected and therefore removed. This can be done either fully automatically or in a semi-automatic fashion following human input (approval).

\section{Summary and conclusions}\label{s:conc}
The aim of this work was to investigate the potential of data-mining messages posted on Twitter to learn about the concerns, practices, and more generally topics of conversation of people interested in the autism spectrum disorder. We approached this problem first by harvesting a large data set of tweets each of which we designated as belonging either to the ASD-related subset of the data corpus or the control subset. Using this data set we conducted a series of experiments which analysed both common and differential characteristics of messages in the two subsets and reported a series of novel results. The most important finding, corroborated by several different experiments, concerns the nature of topics arising in the ASD subset of tweets. We demonstrated that tweets in this subset are very rich in information of potentially high value to public health officials and policy makers, thus motivating further work towards our goal of developing a tool which would be able to monitor automatically the response of the ASD community to various initiatives, legislature, medical advances etc. For example, driven by the finding that a high proportion of the tweets in our corpus contain URL links, one of the directions we wish to pursue in future is that of exploring the value of auxiliary information which can be associated with original messages. We also intend to explore automatic ways of labelling topics semantically using meaningful sentence fragments by back-analysing probabilistically the collected text data for persistent ngrams across documents with shared topics~\cite{Aran2012d}.

\section*{Acknowledgements}
The authors would like to express their sincere gratitude to the anonymous reviewers whose constructive feedback on our original work~\cite{BeykAranPhunVenk+2014} greatly helped shape the present paper. Specifically, we are thankful for their suggestions for additional experiments which were described herein.

\bibliographystyle{unsrt}
%\bibliography{./bib1,./my_bibliography}
\bibliography{./my_bibliography}

\end{document}